\documentclass[useAMS,usenatbib,usegraphicx]{mn2e}

\usepackage{epsf,rotating}
\usepackage{setspace}
\usepackage{subfigure}
\usepackage{mathrsfs}
\usepackage{amssymb,amsmath}

\newcommand{\cmc}{\;{\rm cm}^{-3}}

\newcommand{\Zsolar}{\;{\rm Z}_{\odot}}

\newcommand{\gad}{{\sc Gadget-3}}

\newcommand{\CII}{{\hbox{C\,{\sc ii}}}}

\newcommand{\CIV}{\hbox{C\,{\sc iv}}}
\newcommand{\CV}{\hbox{C\,{\sc v}}}

\newcommand{\SIII}{{\hbox{S\,{\sc iii}}}}

\newcommand{\SiII}{{\hbox{Si\,{\sc ii}}}}

\newcommand{\NV}{\hbox{N\,{\sc v}}}
\newcommand{\OI}{\hbox{O\,{\sc i}}}
\newcommand{\OII}{\hbox{O\,{\sc ii}}}
\newcommand{\OIII}{\hbox{O\,{\sc iii}}}
\newcommand{\OIV}{\hbox{O\,{\sc iv}}}
\newcommand{\OV}{\hbox{O\,{\sc v}}}
\newcommand{\OVI}{\hbox{O\,{\sc vi}}}
\newcommand{\OVII}{\hbox{O\,{\sc vii}}}
\newcommand{\OVIII}{\hbox{O\,{\sc viii}}}
\newcommand{\OIX}{\hbox{O\,{\sc ix}}}
\newcommand{\HI}{{\hbox{H\,{\sc i}}}}
\newcommand{\HII}{{\hbox{H\,{\sc ii}}}}
\newcommand{\HeI}{{\hbox{He\,{\sc i}}}}
\newcommand{\HeII}{{\hbox{He\,{\sc ii}}}}

\newcommand{\NeVIII}{{\hbox{Ne\,{\sc viii}}}}

\newcommand{\MgII}{{\hbox{Mg\,{\sc ii}}}}
\newcommand{\MgX}{{\hbox{Mg\,{\sc x}}}}
\newcommand{\FeI}{\hbox{Fe\,{\sc i}}}
\newcommand{\FeII}{\hbox{Fe\,{\sc ii}}}
\newcommand{\FeXI}{\hbox{Fe\,{\sc xi}}}
\newcommand{\FeXVII}{\hbox{Fe\,{\sc xvii}}}

\newcommand{\nh}{{n_{\rm H}}}
\newcommand{\tcool}{\tau_{\rm cool}}
\newcommand{\tdyn}{\tau_{\rm dyn}}

\newcommand{\tphot}{\tau_{\rm phot}}

\newcommand{\tionOI}{\tau_{\rm ion,\OI}}

\newcommand{\tionOVIII}{\tau_{\rm ion,\OVIII}}
\newcommand{\Jnuunits}{{\rm erg}\,{\rm s}^{-1}\,{\rm cm}^{-2}\,{\rm Hz}^{-1}\,{\rm sr}^{-1}}
\DeclareMathSymbol{\la}{3}{AMSa}{46}
\DeclareMathSymbol{\ga}{3}{AMSa}{38}

\title[Non-equilibrium ionisation and cooling]{Non-Equilibirum ionisation and cooling of metal-enriched gas in the presence of a photo-ionisation background}

\author[B. D. Oppenheimer \& J. Schaye]{
\parbox[t]{\textwidth}{\vspace{-1cm}
Benjamin D. Oppenheimer$^{1,2}$, Joop Schaye$^1$}
\\\\$^1$ Leiden Observatory, Leiden University, PO Box 9513, 2300 RA Leiden, the Netherlands
\\$^2$ CASA, Department of Astrophysical and Planetary Sciences, University of Colorado, Boulder, CO 80309, USA
}

\begin{document}

\pubyear{2013}

\maketitle

\label{firstpage}

\begin{abstract}
Simulations of the formation of galaxies, as well as ionisation models
used to interpret observations of quasar absorption lines, generally
either assume ionisation equilibrium or ignore the presence of the
extra-galactic background (EGB) radiation. We introduce a method to
compute the non-equilibrium ionisation and cooling of diffuse gas
exposed to the EGB. Our method iterates the ionisation states of the
11 elements that dominate the cooling (H, He, C, N, O, Ne, Si, Mg, S,
Ca, \& Fe) and uses tabulated ion-by-ion cooling and photo-heating
efficiencies to update the temperature of the gas. Our reaction
network includes radiative and di-electric recombination, collisional
ionisation, photo-ionisation, Auger ionisation, and charge
transfer. We verify that our method reproduces published results for
collisional equilibrium, collisional non-equilibrium, and
photo-ionised equilibrium. Non-equilibrium effects can become very
important in cooling gas, particularly below $10^6$
K. Photo-ionisation and non-equilibrium effects both tend to boost the
degree of ionisation and to reduce cooling efficiencies. The effect of
the EGB is larger for lower densities (i.e.\ higher ionisation
parameters). Hence, photo-ionisation affects (equilibrium and
non-equilibrium) cooling more under isochoric than under isobaric
conditions. Non-equilibrium effects are smaller in the presence of the
EGB and are thus overestimated when using collisional-only processes.
The inclusion of the EGB alters the observational diagnostics of
diffuse, metal-enriched gas (e.g.\ metal absorption lines probed in
quasar sight lines) even more significantly than the cooling
efficiencies. We argue that the cooling efficiency should be
considered if ionisation models are used to infer physical conditions
from observed line ratios, as the a priori probability of observing
gas is lower if its cooling time is shorter. We provide on-line tables
of ionisation fractions and cooling efficiencies, as well as other
data, for equilibrium and non-equilibrium scenarios, and both with and
without an EGB. Cooling efficiencies and diagnostics of the physical
state of diffuse gas can become highly inaccurate if ionisation
equilibrium is assumed or if the existence of the ionising background
is ignored.
\end{abstract}


\section{Introduction}  

The rates at which diffuse gases cool radiatively are central to
numerous baryonic processes in astrophysics.  In the context of galaxy
formation, how gas cools in the intergalactic medium (IGM) and
galactic haloes critically determines if, how, and when galaxies
receive their gas, which provides the main source of fuel for star
formation.

The ejection of the nucleosynthetic products of star formation (metal
species) into diffuse regions complicates the picture, as these
sub-dominant heavier atomic species are very efficient coolants, owing
to their more numerous electrons and line transitions.  The inclusion
of radiative metal cooling in gas enriched to solar metal abundances
($\Zsolar$) or even $0.1 \Zsolar$ significantly reduces cooling times
and directly leads to more efficient accretion onto galaxies
\citep[e.g.][]{opp06, choi09, sch10, van12, haas12}.

Diffuse regions outside of galaxies are photo-ionised and photo-heated
by the extra-galactic background (EGB) originating from a combination
of UV and X-ray emitting sources including active galactic nuclei
(AGN) and star-forming galaxies \citep[e.g.][]{haa96,haa12}.
Photo-ionisation of hydrogen, helium, and metal species reduces
cooling efficiencies, because fewer bound electrons provide fewer line
transitions for radiative cooling \citep[][hereafter W09]{efs92,
  wie09a}.  W09 showed net cooling efficiencies declined by an order
of magnitude at typical densities and temperatures of the shock-heated
IGM.  Including the effect of photo-ionisation on metal cooling
efficiencies can therefore significantly affect the appearance of metal
absorption in quasar absorption line (QAL) spectra \citep{tep11,
  opp12a} and alter the dynamics of the accreting gas onto galaxies
\citep{smi11, opp12a}.

Most hydrodynamic simulations and semi-analytic models assume that gas
is in ionisation equilibrium, but this assumption can break down if
the ionisation or recombination time of a species is long compared to
other timescales, including the dynamical, Hubble, or cooling
timescales.  For example, \citet[][hereafter GS07]{gna07} calculated
that when the timescale of recombination exceeded that for cooling, a
``recombination lag'' occurs where the gas is over-ionised relative to
the equilibrium case
\citep[e.g.][]{kaf73,sha76,sch93,sut93,cen06b,yos06,vas13}.  By
following the ionisation state of gas without any external radiation
(i.e.\ collisional processes only), they find fewer bound electrons
for metal ion species with long recombination times and that this
results in less efficient net cooling and cooling functions of
temperature with less sharp peaks.  

\citet{vas11} calculated non-equilibrium ionisation and cooling rates
in the presence of a UV/X-ray radiation field, including the
\citet{haa01} field and power-law spectra corresponding to AGN and
stellar clusters.  He demonstrated that photo-ionisation suppressed
cooling by ionising metals to higher states, which combined with the
recombination lag, made photo-ionised non-equilibrium cooling less
efficient than the collisional-only non-equilibrium cases of GS07.

In this paper, we develop a new method to follow non-equilibrium
ionisation and cooling in the presence of a photo-ionising EGB,
focusing on temperatures $\geq 10^4$ K.  By self-consistently tracking
the ionisation state of primordial and metal species, we can
accurately calculate the non-equilibrium cooling rates of
metal-enriched gas.  In \S2, we introduce and test the method that we
have integrated into the \gad~\citep[last described in][]{spr05}
hydrodynamical simulations code for hydrogen, helium, electrons, and
the important metal coolants (carbon, nitrogen, oxygen, neon,
magnesium, silicon, sulphur, calcium, and iron).  To demonstrate the
effects and importance of non-equilibrium processes, we use a
stand-alone, ``single-particle'' version of the code to follow the
ionisation and temperature evolution of a parcel of gas independent of
any hydrodynamics, but still in the presence of an EGB.  We diagnose
how non-equilibrium effects diverge from the equilibrium case for
enriched cooling gases without photo-ionisation in \S3 and with
photo-ionisation in \S4.  Relevant astrophysical applications include
metal-enriched materials ejected into the IGM and the circumgalactic
medium by superwinds, which simulations show regularly re-accrete onto
galaxies at late times \citep[$z\la 1$;][]{opp10, van12}.  We show
that the combination of non-equilibrium effects and photo-ionisation
can retard accretion and significantly alter the observational
diagnostics of cooling gas that may be observed via metal-line
absorption with instruments such as the Cosmic Origins Spectrograph
(COS) on the {\it Hubble Space Telescope}.  Finally, we summarise our
findings in \S5.

We provide on-line tables including cooling and
photo-heating efficiencies per ion, ionisation fractions in
equilibrium and non-equilibrium cases, and the compiled atomic data
used in this work at http://noneq.strw.leidenuniv.nl.  Supplementary
figures are also available at this website.

\section{Method and tests}

Our method explicitly follows the ionisation states of all 11 elements
that contribute significantly to the cooling efficiencies at
temperatures $T\geq 10^4$ K (i.e.\ H, He, C, N, O, Ne, Si, Mg, S, Ca,
\& Fe; W09) as well as the electron density.  These are all the
species tabulated by W09 and used in the OverWhelmingly Large
Simulations (OWLS) project \citep{sch10}.  They account for nearly all
the radiative cooling \citep[W09;][]{ber13}. Our method is meant to be
complimentary to W09, so that we can turn off following the
non-equilibrium ionisation states of any metal species and instead use
the corresponding equilibrium tables.  For example, iron involves
following 27 ionic species, which may exacerbate the memory
requirements when running a large simulation.  The code can be used to
follow only the metal-line coolants that dominate at a given
temperature (C, O, Ne, \& Fe), or it can be used to follow the
ionisation states of species frequently detected in QAL observations
(e.g.\ C, O, Si, \& Mg).  Species including S and Ca, which have many ions
but are not very important for cooling and rarely seen in QAL spectra
may be less useful to follow, so they can be turned off.  We also
build in options to turn off Auger ionisation and charge transfer
reactions, because these calculations can significantly increase
computation time in exchange for mostly minor changes in the
ionisation balance.

Having computed the ionisation states for each species, we can use
ion-by-ion lookup tables as functions of temperature and redshift for
the cooling and photo-heating efficiencies.  Because we explicitly
track ionisation states, the lookup tables are much more compact than
those of W09, who had to track net cooling efficiencies as a function
of density, temperature, redshift, and helium fraction.  Finally, we
consider criteria to cycle and sub-cycle the time integration of
ionisation and cooling.

Throughout, we will assume the gas to be optically thin to both the
EGB and the cooling radiation. Assuming the as to be optically thin is
generally a good approximation for gas with densities $\nh \la 10^{-3}
\cmc$ \citep{rah12}, but we note that for higher densities
self-shielding may reduce the effects of photo-ionisation. We will
ignore sources of ionising radiation other than the EGB, including
self-radiation from shocks and local sources, which is a conservative
assumption when investigating the potential effects of
photo-ionisation. We will also assume that the electron and ion
temperatures are equal. Although the electron temperature may
temporarily lag that of the ions in shock-heated gas, supernova
remnants suggest that plasma waves quickly equilibrate the different
temperatures \citep[e.g.][and references therein]{ber08}.

We first discuss the method and tests of following the
ionisation of the gas, then the cooling and photo-heating method and
tests, and finally how we cycle and subcycle during timesteps.  

\subsection{Ionisation} \label{sec:ionmethod}

We consider the processes of radiative and di-electric recombination,
collisional ionisation, photo-ionisation by an EGB, Auger ionisation,
and charge transfer in our calculations of the ionisation balance.
For an ionisation state $i$ of element $x$, the time dependent
evolution (omitting Auger ionisation and charge transfer for
simplicity) of the number density, $n$, of each ion species, $x_i$, is
given by

\begin{equation} \label{eqn:ionstate}
\begin{split}
\frac{dn_{x_i}}{dt}& = n_{x_{i+1}} \alpha_{x_{i+1}} n_e + n_{x_{i-1}}(\beta_{x_{i-1}} n_e + \Gamma_{x_{i-1}})\\
& - n_{x_i} ((\alpha_{x_i}+\beta_{x_i}) n_e +\Gamma_{x_i}),
\end{split}
\end{equation}

\noindent where $n_e$ is the free electron density (cm$^{-3}$),
$\alpha_{x_i}$ is the total recombination rate coefficient (radiative
plus dielectronic, cm$^3$ s$^{-1}$), $\beta_{x_i}$ is the collisional
ionisation rate coefficient (cm$^3$ s$^{-1}$), and $\Gamma_{x_i}$ is
the photo-ionisation rate (s$^{-1}$) for the given ionisation state,
where

\begin{equation} \label{eqn:gamma}
\Gamma_{x_i} = \int^{\infty}_{\nu_{0,x_i}} \frac{4 \pi J_{\nu}}{h \nu} \sigma_{x_i}(\nu) d\nu,
\end{equation}

\noindent where $\nu$ is frequency, $\nu_{0,x_i}$ is the ionisation
frequency, $J_{\nu}$ is the EGB radiation field ($\Jnuunits$),
$\sigma_{x_i}(\nu)$ is photo-ionisation cross-section, and $h$ is the
Planck constant.

We use atomic data similar to that used by CLOUDY
ver. 10.00\footnote{http://www.nublado.org/} \citep[last described
  in][]{fer98} where reasonably possible.  For radiative and
dielectronic recombination coefficients, we use the \citet{bad06} and
\citet{bad03} fits, respectively, if available for our species.
Otherwise, we use the \citet{ver96a} fits for radiative recombination
if available, and finally recombination data compiled by
D. Verner\footnote{http://www.pa.uky.edu/$\sim$verner/rec.html} from
\citet{ald73, shu82, arn85}.  We assume case A recombination.  Fits to
collisional ionisation rate coefficients are from \citet{vor97}.  We
tabulate these coefficients between log[$T$ (K)]$=2.0-9.0$ at 0.04 dex
intervals resulting in 176 temperature bins, which is the same spacing
as used by W09.

We perform the integral in Equation \ref{eqn:gamma} using the same
$\sigma_{x_i}(\nu)$ as used in CLOUDY ver. 10.00 \citep{ver95, ver96}
for the appropriate EGB.  In this work, we use the \citet[][hereafter
  HM01]{haa01} EGB as our fiducial background since this results in
good agreement with the observed column density distribution of $\HI$
\citep{dav10, alt11} and metal-line ratios \citep{agu08}.  This EGB
includes contributions from quasars and galaxies and is tabulated at
50 redshifts between $z=0-9$ sampling at constant log(1+$z$)
intervals, resulting in a lookup table of $\Gamma_{x_i}$s as a
function of redshift corresponding to each ion containing electrons.
We also tabulate photo-ionisation rates for the \citet{fau09} and
\citet{haa12} and make these available on our website.

Auger ionisation involves the ejection of multiple electrons following
the ionisation of an inner shell electron.  The Auger reaction is
stimulated by an energetic photon, $\gamma$, such that when element A,
initially ionised to charge $+n$, interacts with $\gamma$, $m$
electrons are released.  The rate equation is then A$^{+n}$ + $\gamma$
$\rightarrow$ A$^{+n+m}$ + $m$ e$^-$, where $m> 1$.  We use the
electron vacancy distribution probabilities calculated by
\citet{kaa93}, and re-tabulated for CLOUDY as probabilities of
ionising multiple numbers of electrons, all the way up to 10 electrons
for the case of $\FeI$ to $\FeXI$.  Auger ionisation is significant
for photo-ionised oxygen species, because 2 electron losses are common
for inner shell ionisations of $\OI$ through $\OV$.  We include the
option to turn Auger ionisation off, because it slows the calculation,
or to lower $m$ to values as low as two, if one does not want to
follow the $m>2$ reactions that are mostly limited to heavier species
like Fe.

Charge transfer reactions encompass two-body reactions where neutral H
or He exchange an electron with a heavier ion species leading to
recombination of the heavier ion.  For hydrogen reacting with element
A ionised to charge $+n$, the reaction is H$^0$ + A$^{+n}$
$\rightarrow$ H$^+$ + A$^{+n-1}$.  The most significant reaction is
the recombination of $\OII$ to $\OI$ owing to $\HI$, which can enhance
$\OI$ levels.  The reverse reactions of singly ionised H and He with a
heavier species ionises the heavier species such that for hydrogen,
H$^+$ + A$^{+n-1}$ $\rightarrow$ H$^0$ + A$^{+n}$.  CLOUDY uses rates
from various references \citep[e.g.][]{kin96} to calculate the four
varieties of charge transfer reactions: recombination and ionisation
with H and He.  We have outputted and tabulated these rates for our
176 temperatures from CLOUDY for all relevant metal ions with H and
He, as well as charge transfer between H and He.

We use the Sundials {\tt
  CVODE}\footnote{https://computation.llnl.gov/casc/sundials/main.html}
solver to integrate the ionisation states over a timestep $dt$ to
solve for each $x_i$.  We use the backward difference formula (BDF)
method and Newton iteration in {\tt CVODE}, given the set of rate
equations.  For an atomic species with atomic number $Z$, there are
$Z+1$ versions of Equation \ref{eqn:ionstate} to solve for the $Z+1$
ions that exist; hence the number of equations equals the number of
unknowns.  In practice we follow all $Z+1$ ions, even though an
independent constraint is available, namely that the sum of all ion
fractions equals one, which allows one to follow only $Z$ states.  We
calculate $\sum\limits_i f_{x_{i}}$ for each element, where
$f_{x_{i}}$s are ion fractions, as an independent check to see how
well the BDF method is tracking the states, which depends on the
accuracy that one chooses for the calculations using the relative and
absolute tolerance parameters described in the {\tt CVODE} manual.
Briefly, for the calculations in this paper we set absolute tolerance
vector to $10^{-5}\times f_{x_i}$ for metal species and $10^{-7}\times
f_{x_i}$ for hydrogen and helium.  The relative tolerances are
$10^{-5}$ and are unlikely to play a role, since the absolute
tolerance vectors are equal or tighter, plus they act as relative
tolerances since they are multiplied by the ionisation fraction.
However, for the version integrated into \gad, we will choose looser
tolerances according to the computational resources.  If
$\sum\limits_i f_{x_i}$ strays more than 1\% from unity, we
renormalize the ionisation fractions.  However, such deviations occur
only in extreme situations and are especially rare given our low
tolerances.  Deviations from unity indicate that the timestep given to
the {\tt CVODE} solver is too large, which mostly affects the species
with the most ions (e.g. Fe).

We combine all atomic species into one {\tt CVODE} calculation, which
includes $Z+1$ ions for each atomic species, as well as $n_e$, which
we also follow in the solver.  Including all eleven atomic species, we
follow 134 rate equations.  We include the option to turn off the rate
equation for $n_e$ if Auger ionisation and charge transfer are also
turned off, because $n_e$ depends on as many as 133 ions and ruins the
symmetry of the Jacobian matrix (which Auger ionisation and charge
transfer also break), causing the calculation to scale non-linearly
with the number of rate equations.  As most of the IGM is highly
ionised and $n_e$ changes very slowly outside of the $\HI$ and $\HeII$
reionisation epochs, we build in the option to update $n_e$ only after
each CVODE iteration.  For the calculations in this paper we follow
everything self-consistently.  Similarly, we leave Auger ionisation
and charge transfer on in our calculations, but we generally recommend
turning these off in a hydro simulation owing to longer computation
times from non-symmetrical entries in the Jacobian.

\subsubsection{Ionisation test cases}

\begin{figure*}
  \subfigure{\setlength{\epsfxsize}{0.32\textwidth}\epsfbox{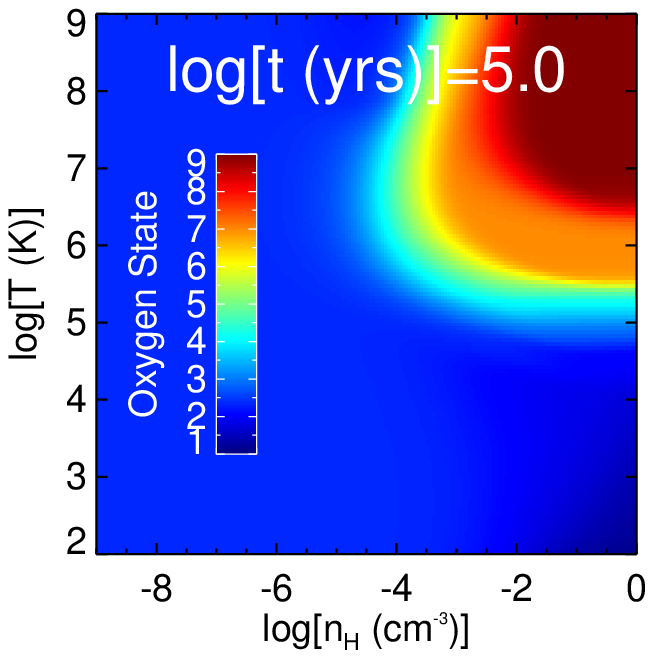}}
  \subfigure{\setlength{\epsfxsize}{0.32\textwidth}\epsfbox{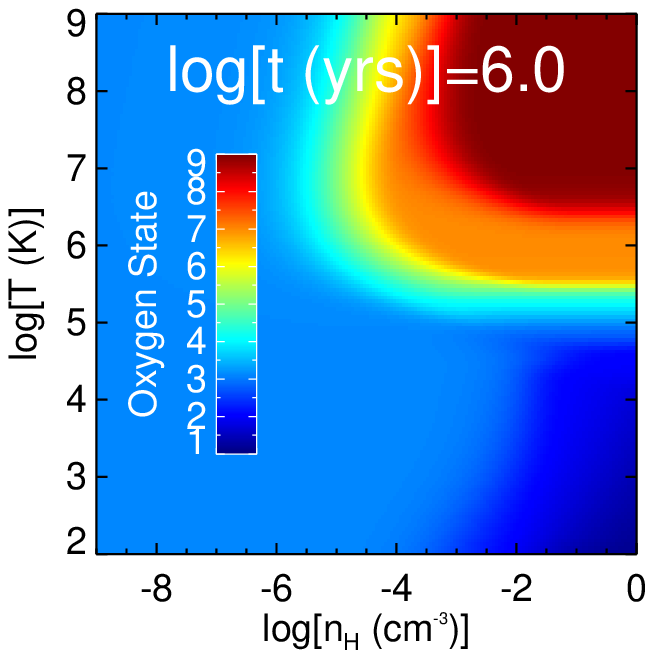}}
  \subfigure{\setlength{\epsfxsize}{0.32\textwidth}\epsfbox{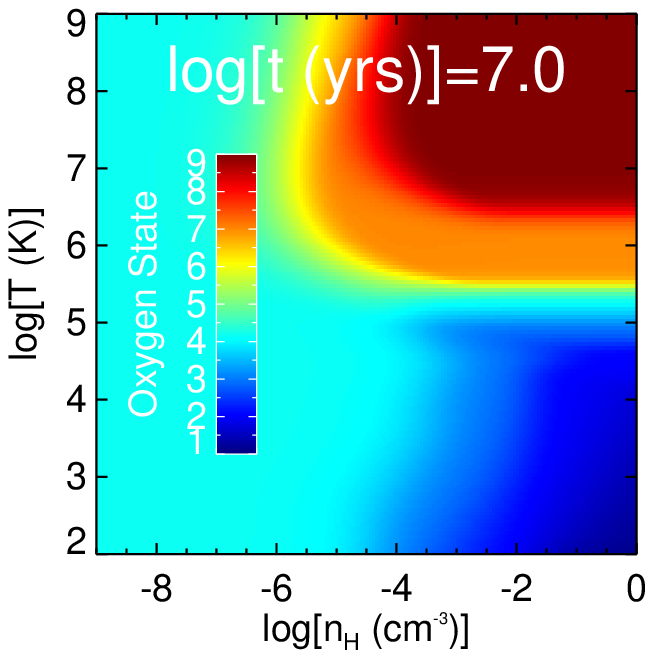}}
  \subfigure{\setlength{\epsfxsize}{0.32\textwidth}\epsfbox{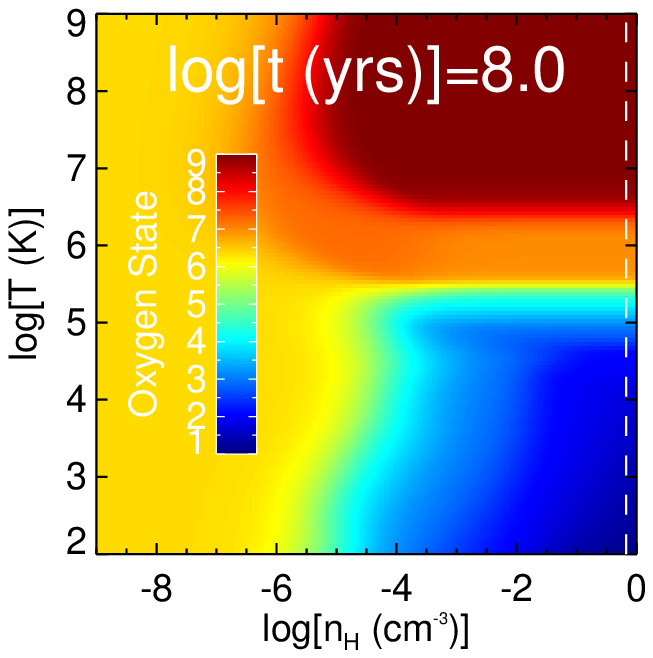}}
  \subfigure{\setlength{\epsfxsize}{0.32\textwidth}\epsfbox{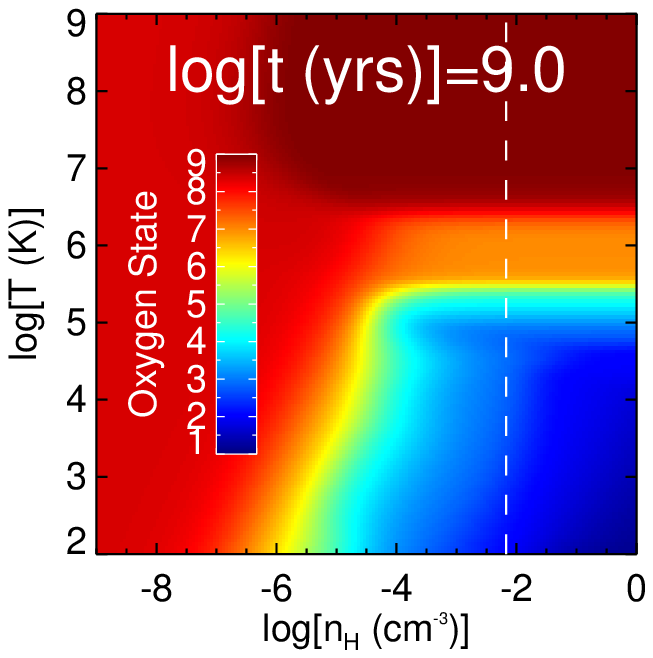}}
  \subfigure{\setlength{\epsfxsize}{0.32\textwidth}\epsfbox{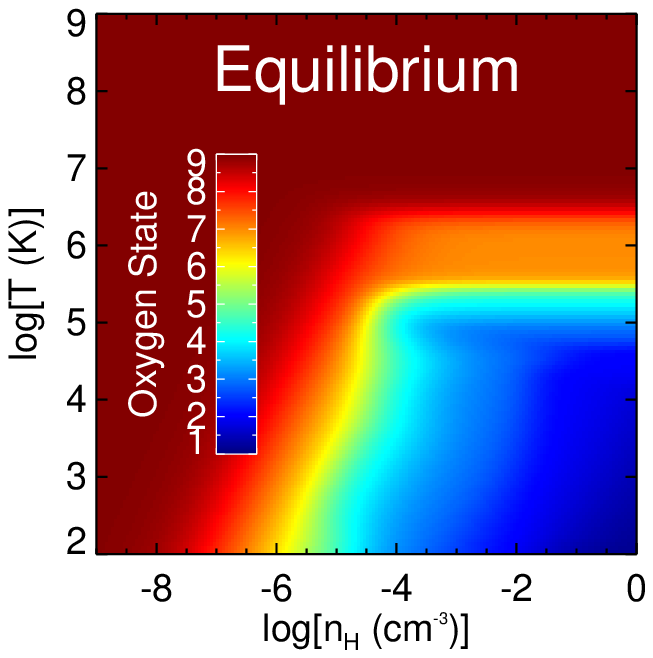}}
\caption{The evolution of non-equilibrium oxygen ionisation fractions
  starting from a neutral state at $t=0$ for a range of densities and
  temperatures using the $z=1$ HM01 radiation field.  The colour
  indicates the average ionisation fraction, from $\OI$ (dark blue) to
  $\OIX$ (dark red).  The first five panels correspond to the five
  distinct times indicated and the bottom right panel corresponds to
  the equilibrium case, which is reached by 10 Gyr for all densities.
  The vertical, dashed white line indicates where the dynamical time
  equals $t$, which only corresponds to densities in the displayed
  range at 100 Myr and 1 Gyr.  This implies the area to the left of
  this line (the entire phase spaces at $\leq 10$ Myr) would not be
  dynamically disrupted.  Collisional ionisation timescales are
  shorter at high densities, achieving equilibrium values earlier.
  Photo-ionisation timescales are independent of density and
  equilibrium has not yet been achieved by 1 Gyr at the lowest
  densities where photo-ionisation dominates.  For movies of the
  evolution, please visit http://noneq.strw.leidenuniv.nl.}
\label{fig:Oionrhotevol}
\end{figure*}

We demonstrate the capabilities of the ionisation solver by
considering the evolution of initially neutral, solar abundance gas at
a large range of densities and temperatures corresponding to diffuse
gas regions.  Figure \ref{fig:Oionrhotevol} shows the ionisation state
of oxygen in density-temperature phase space at five distinct times
($t= 0.1$ Myr, 1 Myr, 10 Myr, 100 Myr, \& 1 Gyr) after the neutral
initial conditions, and also in equilibrium, which is nearly
equivalent to the case after 10 Gyr.\footnote{For movies showing time
  evolution of the ionisation states, and also for other metal
  species, please visit http://noneq.strw.leidenuniv.nl.} In this
example, we assume the $z=1$ HM01 EGB and neglect redshift evolution.
We apply no cooling or photo-heating in this example, allowing us to
focus on the ionisation behaviour by leaving the density and
temperature constant.  At $t=0$ (not shown), the entire phase space
would be dark blue, corresponding to $\OI$, while after 1 Myr, the
highest densities and temperatures (regions that would typically be
defined as the intracluster medium (ICM), $\nh\ga 10^{-3} \cmc$,
$T>10^{7}$ K) have achieved complete ionisation to $\OIX$,
corresponding to dark red.

The evolution of the oxygen ionisation state over the next Gyr
demonstrates how significant ionisation timescales lead to
non-equilibrium ionisation behaviour.  The ionisation timescale for a
given ion $i$ is
\begin{equation}
\tau_{{\rm ion}, {x_i}} = \frac{1}{\beta_{x_i} n_e+\Gamma_{x_i}}.
\end{equation}
\noindent The collisional ionisation timescale, $\tau_{{\rm
    coll},{x_i}}= 1/(\beta_{x_i} n_e)$, depends inversely on density;
hence the highest densities achieve equilibrium in less than a Myr.
This includes gas collisionally ionised to $\OIX$, which depends on
eight ionisation timescales, $\tionOI$ through $\tionOVIII$, all of
which are $\ll 1$ Myr at the highest densities shown in Figure
\ref{fig:Oionrhotevol}.  For each 1 dex increase in $t$, the density
achieving collisional ionisation equilibrium (CIE), defined as where
collisional ionisation exactly balances recombination, decreases by 1
dex.  In CIE, the ionisation state depends only on $T$, which is why
there is no density dependence at the highest densities where
photo-ionisation by the uniform HM01 EGB is unimportant ($\beta_{x_i}
n_e \gg \Gamma_{x_i}$).  In CIE the $\OVI$ fraction peaks at $T\approx
10^{5.5}$ K, corresponding to the horizontal yellow band indicating
this state on the right-hand side of each panel in Figure
\ref{fig:Oionrhotevol}.

The photo-ionisation timescales, $\tau_{{\rm phot},{x_i}}=
1/\Gamma_{x_i}$, are independent of density for a uniform EGB.  At low
densities where $\tau_{{\rm phot},{x_i}}\ll \tau_{{\rm coll},{x_i}}$,
the ionisation state depends on the series of $\tphot$s, which
increase with ionisation state.  After 100 Myr the dominant ionisation
state at low density ($\nh\la 10^{-6} \cmc$) is $\OVI$ and after 1 Gyr
it is $\OVIII$.  The lowest densities, corresponding to the centres of
voids at $z=1$ ($\nh \la 10^{-7} \cmc$), take nearly $10$ Gyr
(assuming the $z=1$ HM01 field) to achieve their equilibrium
ionisation states, which are dominated by $\OIX$ for all temperatures.

The dashed vertical white line, indicates the density for which the
dynamical timescale,
\begin{equation}
\tdyn = 1/\sqrt{G \rho}, 
\end{equation}
\noindent equals $t$, where $G$ is the gravitational constant and
$\rho$ is the gas mass density.  $\tdyn$ is long compared to $t$ to
the left of this line.  The line is beyond the right boundary of the
$\leq 10$ Myr panels, meaning that the gas exhibiting non-equilibrium
ionisation states will, at these times, not be disrupted by dynamical
processes.  An astrophysically relevant case may be the injection of
enriched, low-ionisation wind material into the IGM, where an
``ionisation lag'' leaves the gas under-ionised relative to
equilibrium values.  \citet{cen06b} found enriched gas was
under-ionised in their cosmological simulation that explicitly
followed the oxygen states $\OV-\OIX$, which was reflected in fewer
predicted $\OVII$ and $\OVIII$ absorption line detections in their
mock AGN X-ray spectra probing the IGM.

\begin{figure}
  \subfigure{\setlength{\epsfxsize}{0.49\textwidth}\epsfbox{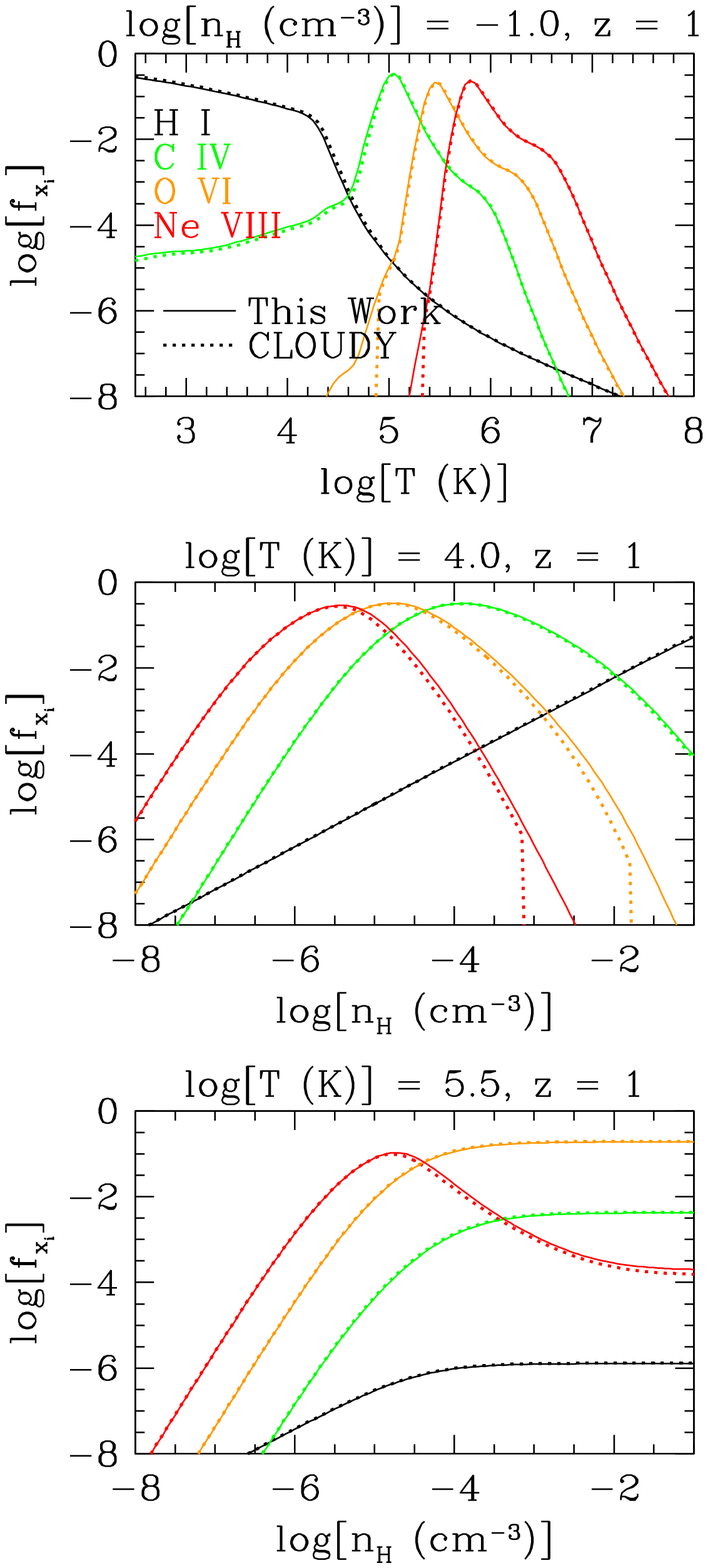}}
  \caption{Comparisons of equilibrium ionisation fractions for $\HI$,
    $\CIV$, $\OVI$, and $\NeVIII$ predicted by our code and by CLOUDY
    using the $z=1$ HM01 radiation field.  The top panel shows ion
    fractions as a function of temperature at the constant density
    $\nh=0.1 \cmc$, for which collisional ionisation dominates in most
    species.  The bottom two panels show ion fractions as a function
    of density at $T=10^{4.0}$ and $10^{5.5}$ K, respectively.
    Photo-ionisation dominates the middle panel while the bottom panel
    tests the transition between the photo- and collisionally ionised
    regimes. The agreement with CLOUDY is excellent, although there is
    deviation at some very low metal ion fractions where CLOUDY does
    not accurately track these very low abundances.}
  \label{fig:ioncloudycheck}
\end{figure}

We provide a consistency check for the predicted ionisation fractions
of our code by comparing to CLOUDY ver. 10.00 calculations in Figure
\ref{fig:ioncloudycheck} for gas in equilibrium, which we obtain for
our code by integrating for at least 100$\times$ any recombination or
ionisation timescale.

The top panel shows the equilibrium ionisation fractions for $\HI$,
$\CIV$, $\OVI$, and $\NeVIII$ as a function of temperature for a
density where collisional ionisation dominates, $\nh= 0.1 \cmc$.
The dotted lines are from CLOUDY-generated ionisation lookup tables at
$z=1$ assuming equilibrium.  Our code produces the solid lines that
reproduce the CLOUDY CIE behaviour of the four ions, which depend on
the balance of recombination and collisional ionisation.

The bottom two panels of Figure \ref{fig:ioncloudycheck} demonstrate
the impact of photo-ionisation as a function of density for $T=10^4$ K
(middle) and $10^{5.5}$ K (bottom).  Photo-ionisation dominates in the
$T=10^4$ K case and the two codes agree well.  For $\HI$, $n_{\rm HI}
\propto \nh$ because recombination ($\propto n_e$) balances
photo-ionisation (independent of $n_e$) for this two-state species in
equilibrium.  In contrast, the lithium-like metal ions exhibit a peak
ionisation fraction corresponding to a preferred density.  As the
atomic number increases, the energy required for ionisation increases,
and the number of photons capable of ionising decreases for realistic
spectra.  Hence, the density at which the equilibrium ion fraction
peaks declines with increasing atomic weight.

The $T=10^{5.5}$ K case shows that our code reproduces the transition
from photo-ionisation to collisional ionisation at higher densities
for all species.  We choose this temperature specifically because
$\OVI$ achieves its CIE peak at this temperature.  $f_{\rm HI}$
asymptotes to $10^{-6}$ in the collisional ionisation regime for this
temperature.  For further comparison tests between the equilibrium
abundances predicted by our code and CLOUDY, we forward the reader to
the Appendix and additionally our website--
http://noneq.strw.leidenuniv.nl.  We also demonstrate the effects of
Auger ionisation and charge transfer in the Appendix.  The examples in
this Section have both of these processes turned on.

While these comparisons demonstrate that our reaction network contains
all the important processes and that our atomic data is close to that
used by CLOUDY, they do not test the accuracy of time integration.  We
will test the time-dependent non-equilibrium behaviour in
\S\ref{sec:coll} when we compare the results of our method to GS07.

\subsection{Radiative cooling and photo-heating}

The net volumetric cooling rate ($\Lambda_{\rm net}$) is the
difference between cooling and photo-heating rates.  W09 tabulated net
cooling tables under the assumption of ionisation equilibrium as
functions of temperature, density, redshift, and helium fraction for 9
metals and primordial species separately.  Here we make use of
ion-by-ion cooling and photo-heating tables that are more compact and
flexible, which can be combined into net cooling functions for both
equilibrium and non-equilibrium calculations.  Our ion-by-ion net
cooling method uses the ion fraction calculated using the {\tt CVODE}
solver.  We calculate net cooling as the difference between cooling
rates ($\Lambda'_{x_i} n_{x_i} n_e$) and photo-heating rates
($\epsilon_{x_i} n_{x_i}$), where the resulting net cooling function
for an ion $x_i$ is
\begin{equation}
\Lambda_{{\rm net},{x_i}}(T,z,n_{x_i},n_{e}) = \Lambda'_{x_i}(T) n_{x_i} n_{e} - \epsilon_{x_i}(z) n_{x_i}
\end{equation}
\noindent The units of $\Lambda'_{x_i}$, ion-by-ion cooling
efficiencies, are erg cm$^{3}$ s$^{-1}$ where we divide the volumetric
cooling rate for an ion by $n_{x_i} n_e$ to make $\Lambda'$
independent of density (at least in the regime where electron-ion
collisions dominate the cooling rate). Photo-heating is calculated in
an analogous manner as photo-ionisation. The photo-heating efficiency
for species $x_i$ defined as
\begin{equation} \label{eqn:eps}
\epsilon_{x_i} = \int^{\infty}_{\nu_{0,x_i}} \frac{4 \pi J_{\nu}}{h \nu} \sigma_{x_i}(\nu) h (\nu-\nu_{0,x_i})d\nu, 
\end{equation}
\noindent and hence has unit ergs s$^{-1}$.  We calculate
$\epsilon_{x_i}$ for our 133 $x_i$ species in units of erg s$^{-1}$
for every ion given the HM01 EGB at the 50 redshifts between $z=0$ and
$9$, tabulating it in a table analogous to that of the
photo-ionisation rates.

We have used our own method to calculate CLOUDY cooling efficiencies
ion-by-ion ($\Lambda'_{x_i}$), which was developed independently from
\citet{gna12}, who also tabulate ion-by-ion cooling functions using
the same version (ver. 10.00) of CLOUDY.  We run CLOUDY to calculate
individual ion cooling efficiencies at 176 temperature values between
log[$T$ (K)]$=2.0$ and $9.0$.  The end result is a set of cooling
tables that include the effects of radiative and dielectronic
recombination, collisional ionisation and excitation, and
Bremsstrahlung.  We have compared our ion cooling functions to those of
\citet{gna12} and find that our efficiencies are indistinguishable from
theirs.\footnote{Further comparison figures are available on our
  website.}

In detail, we tabulate ion-by-ion cooling functions by running CLOUDY
with the metal ion fraction we are calculating set to unity.  We
assume solar abundances as was done in W09.  We force hydrogen and
helium to be fully ionised, resulting in $n_e=1.2 \nh$ from these
species alone.  We then run CLOUDY again with the metal being
calculated turned off, and difference the two sets of cooling
efficiencies to obtain $\Lambda'_{x_i}$, which we tabulate for every
metal ion in units of erg cm$^{3}$ s$^{-1}$, because we divide out
$n_{x_{i}} n_e$ from the volumetric cooling rates to maintain density
invariance.  In practice the difference in $n_e$ with the metal ion
turned on and off is very small, since metal species do not contribute
many electrons compared with H and He unless the metallicity is much
higher than solar.  The calculation for primordial ions is similar,
except that the electron density will not default to $n_e=1.2 \nh$ in
the cases of $\HI$, $\HeI$, and $\HeII$, because a significant
fraction of the electrons can become bound.  Therefore, we force $n_e$
in our CLOUDY scripts to add in otherwise bound electrons for these
primordial species, resulting in $n_e=1.2 \cmc$.  Our method differs
from \citet{gna12}, who essentially make a single ion plasma to
calculate cooling efficiencies.  A strategy similar to ours was used
to create the W09 tables except that W09 did not force ion states
since its cooling was calculated assuming ionisation equilibrium.

The total summed net radiative cooling rate per unit volume of all
species is
\begin{equation}
\Lambda_{\rm net} = \sum_x \sum_i \Lambda_{{\rm net},{x_i}} + \Lambda_{\rm Comp},
\end{equation}
where inverse Compton cooling off the cosmic microwave background
contribute $\Lambda_{\rm Comp}=5.64\times 10^{-36}$ erg K$^{-1}$
s$^{-1}$ $(T-T_{\rm CMB,0}\times (1+z)) (1+z)^4 n_e$ , using $T_{\rm CMB,
  0}=2.728$ K \citep{ost06}.

The temperature evolution is calculated using 

\begin{equation} \label{equ:dTdt}
\frac{dT}{dt} = -\frac{\Lambda_{\rm net}}{(3/2+s) n k} + \frac{T}{\mu}\frac{d\mu}{dt},
\end{equation}

\noindent where $n$, the total number density of the gas, is
$\rho/(\mu m_{\rm H})$ and $\mu$, the mean particle mass, is defined
as

\begin{equation}
\mu = \frac{\rho}{n m_{\rm H}}
 = \frac{\rho}{n_e+\sum\limits_i n_{x_i}}, 
\end{equation}

\noindent where $k$ is the Boltzmann constant, and $s$ is 0 for
isochoric (i.e. constant density) cooling and 1 for isobaric
(i.e. constant pressure) cooling.  The isobaric case cools more slowly
owing to performing ``PdV'' work, and we refer the reader to \S2.2 and
\S3 of GS07 for a more complete explanation of isochoric and isobaric
cooling.  We mainly focus on isochoric examples, because this is what
is applicable to integrating cooling during a hydrodynamic timestep in
a simulation.  Lastly, the second term in Equation \ref{equ:dTdt}
accounts for the change in internal energy owing to the change in the
number of particles sharing the internal energy in equipartition.

We define the cooling timescale as

\begin{equation}
\tcool = \frac{(3/2+s) n k T}{\Lambda_{\rm net}}. 
\end{equation}

\noindent and provide this timescale with data included on our
website.

We will continue to update our ion-by-ion cooling data at our website
http://noneq.strw.leidenuniv.nl.  We note the recent improvements in
atomic data used to calculate radiative cooling by \citet{lyk13}.  Our
website will be updated to provide new cooling and ionisation
coefficients from the most recently released official version of
CLOUDY.

\subsubsection{Cooling test cases}

\begin{figure}
  \subfigure{\setlength{\epsfxsize}{0.45\textwidth}\epsfbox{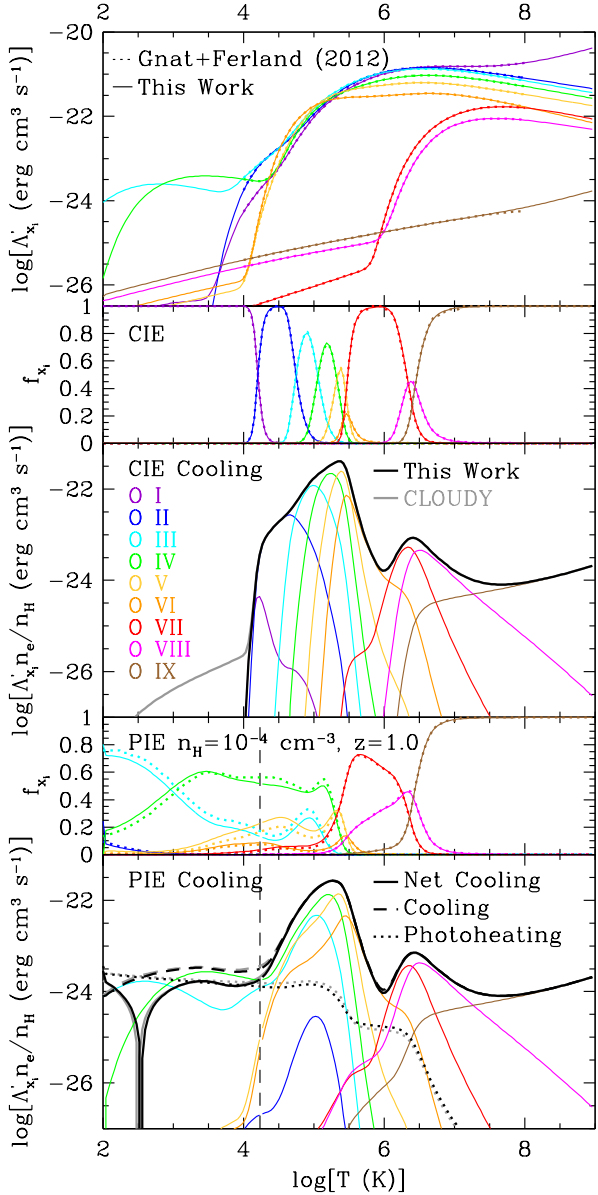}}
\caption[]{The top panel shows the cooling efficiencies of individual
  oxygen ions at solar abundance calculated by us using CLOUDY from
  $T=10^2-10^9$ K (solid lines) compared to efficiencies calculated by
  \citet{gna12} using CLOUDY (dotted lines).  The second panel
  compares the collisional ionisation equilibrium (CIE) oxygen
  fractions we calculate using our reaction network (solid lines) to
  those predicted by CLOUDY (dotted lines).  The integrated oxygen CIE
  cooling efficiency, plotted as $\Lambda'_{x_i} n_e/\nh$ (solid black
  line) in the third panel, is the sum of ion fractions multiplied by
  the ion cooling efficiencies (coloured lines), and agrees well with
  the CLOUDY case (grey line), except below $T=10^4$ K where our
  method loses accuracy due to the presence of cooling channels other
  than electron-ion collisions.  The photo-ionisation equilibrium
  (PIE) oxygen ionisation fractions and summed cooling efficiency
  appear in the bottom two panels, assuming $n_{\rm H}=10^{-4} \cmc$
  and the HM01 EGB at $z=1$.  Cooling, photo-heating
  ($\epsilon_{x_i}/\nh$), and net cooling ($\Lambda_{\rm net}/\nh^2$)
  are indicated in the bottom panel by dashed, dotted, and solid lines
  respectively for our code (black) and CLOUDY (grey).  The
  equilibrium temperature of the solar abundance gas is indicated by
  the vertical dashed line.}
\label{fig:coolcomp}
\end{figure}

The top panel of Figure \ref{fig:coolcomp} shows the cooling functions
($\Lambda'_{x_i}$) for individual oxygen ions and compares them to
those of \citet{gna12}.  We check that the two methods agree after
scaling to CLOUDY default solar abundances.\footnote{These abundances,
  also used in W09, are $n_{\rm He}/n_{\rm H}=0.1$, $n_{\rm C}/n_{\rm
    H}=10^{-3.61}$, $n_{\rm N}/n_{\rm H}=10^{-4.07}$, $n_{\rm
    O}/n_{\rm H}=10^{-3.31}$, $n_{\rm Ne}/n_{\rm H}=10^{-4.00}$,
  $n_{\rm Mg}/n_{\rm H}=10^{-4.46}$, $n_{\rm Si}/n_{\rm
    H}=10^{-4.46}$, $n_{\rm S}/n_{\rm H}=10^{-4.74}$, $n_{\rm
    Ca}/n_{\rm H}=10^{-5.64}$, and $n_{\rm Fe}/n_{\rm H}=10^{-4.55}$.}
The total volumetric cooling rate due to element $x$ is $\sum\limits_i
\Lambda'_{x_i} n_{x} n_{e} f_{x_{i}}$.  Our CIE oxygen ionisation
fractions (solid lines) are compared to the CLOUDY fractions (dotted
lines) in the second panel, demonstrating excellent agreement.  The
third panel shows the summed oxygen CIE cooling curve (thick black
line) multiplied by $n_e/n_{\rm H}$ to plot cooling efficiency,
combined with the contribution of each ion species (coloured lines).
We compare this to the analogous CLOUDY calculation (grey solid line),
which agrees almost perfectly above $T=10^4$ K and is virtually
indistinguishable from our summed cooling curve. The cooling rate
drops precipitously below $T\sim 10^4$ K, because in CIE $n_e$
declines steeply as hydrogen recombines.  However, the CLOUDY
calculation diverges and has much higher cooling owing to collisions
other than electron-ion collisions, in this case $\HI$-$\OI$
collisions.  Our ion-by-ion cooling method fails in this regime for
CIE, but we can still safely apply our method to photo-ionised cases
below $T=10^4$ K where electron-ion collisions dominate cooling.

In the presence of a photo-ionisation background three things happen.
First, ionisation removes electrons meaning higher ions exist at lower
temperatures.  This alone reduces cooling efficiencies in most cases,
since higher ions usually have lower $\Lambda'_{x_i}$ at a given
temperature, because fewer radiative transitions exist; however there
are exceptions.  The fourth panel in Figure \ref{fig:coolcomp} shows
the photo-ionisation equilibrium (PIE) ion fractions of oxygen for our
code (solid lines) and the CLOUDY calculation (dotted lines).  Here we
use a density $n_{\rm H}=10^{-4} \cmc$ and the HM01 EGB at $z=1$. For
states $T\ga 10^5$ K the agreement is excellent while there are some
small, but noticeable deviations at lower temperatures for $\OIII$,
$\OIV$, and $\OV$.  The PIE cooling efficiencies in the bottom panel
show very good agreement.  Our investigation in Appendix
\ref{sec:ioncompare} shows that low ions often agree less well with
CLOUDY calculations than high ions.

Second, $n_e$ does not decline at $T<10^4$ K, because photo-ionisation
keeps most hydrogen ionised.  Hence, for $T<10^4$ K the summed cooling
of oxygen in the bottom panel (dashed thick black line) is actually
much {\it higher} for PIE than for CIE owing to the much greater
electron density.  We provide tables down to $T=10^2$ K, because
electron-ion collisions dominate the cooling in this regime given a
realistic background.

Third, the total volumetric photo-heating of oxygen, $\sum\limits_i
\epsilon_{x_i} n_{x} f_{x_{i}}$, represented by the dotted thick black
line in the bottom panel, adds energy to the gas.  We do not show the
contributions of individual species to the photo-heating as we do for
the cooling.  We plot net cooling, which is defined here as cooling
minus heating, as the solid thick black line.  For $T<10^4$ K even the
net cooling due to oxygen is greater in PIE than in CIE!  The point
where cooling equals heating is $10^{2.50}$ K for oxygen.  However,
the thermal equilibrium point for this entire solar abundance gas is
at $10^{4.23}$ K, mainly owing to hydrogen, and is indicated by the
dashed vertical line in the bottom two panels.  We compare to the
CLOUDY values for cooling, heating, and net cooling with thick grey
lines, finding very good agreement.

\subsection{Cycling and sub-cycling}

Given a timestep, which we will call the cycle timestep, $dt$, which
may be the Courant timestep in a hydrodynamic code, we determine if
the cooling rate needs to be updated on a shorter, sub-cycle timestep
by determining if $\delta u/u > \xi_{\rm cool}$ where $u$ is the
internal energy density, $\delta u\equiv \Lambda_{\rm net}\times dt$
is the change in this energy, and $\xi_{\rm cool}$ is the fractional
cooling tolerance.  We then update the ionisation state explicitly on
either the cycle timestep if $\delta u/u \leq \xi_{\rm cool}$, or on
the sub-cycle timestep defined by $u/\Lambda_{\rm net}\times\xi_{\rm
cool}$.  For the calculations presented here, we use $\xi_{\rm
cool}=0.01$, but we find that running with 0.05 gives acceptable
accuracy for much less computation time.

If we need to sub-cycle, we iterate cooling and ionisation, where the
latter is integrated using CVODE as discussed in
\S\ref{sec:ionmethod}, explicitly until either i) the cycle timestep
is reached, or ii) the equilibrium temperature is achieved by noting
that the sign of $\Lambda_{\rm net}$ reverses.  If a sign reversal
occurs, we backtrack to the equilibrium temperature where
$\Lambda_{\rm net}=0$, since further subcycling is unnecessary and
computationally expensive.  This is mainly a cosmetic choice as the
internal energy will wiggle around the equilibrium temperature with a
tolerance according to $\xi_{\rm cool}$ between cycles if the particle
remains at the equilibrium temperature.  The ionisation state will
also wiggle, because the ionisation timescales are typically very
short; hence our extra timestep attempts to prevent this unsightly
wiggling.

Finally, it should be noted that the ionisation sub-cycle timestep is
determined by cooling.  The reverse, i.e.\ the cooling timestep being
sub-cycled according to changes in the ionisation state, is not
possible in our explicit method described thus far.  In most cases
applicable to diffuse gas, the cooling efficiency is not dramatically
altered by a change in the ionisation state, but there are some
notable exceptions.  The most obvious example is reionisation, where
the ionisation timescale is short and the cooling characteristics
change dramatically from neutral to ionised.  Another case is that of
a nearby AGN suddenly turning on or off, changing the ionisation state
on a timescale much shorter than the cooling time.  We will explore
the latter case in Oppenheimer \& Schaye (in preparation), where we
will also discuss how we address the sub-cycling.  For now, we state
that our current method is meant to follow cooling in the presence of
a slowly changing EGB.

\section{Collisionally ionised non-equilibrium metal-enriched gases} \label{sec:coll}

\begin{figure*}
  \subfigure{\setlength{\epsfxsize}{0.98\textwidth}\epsfbox{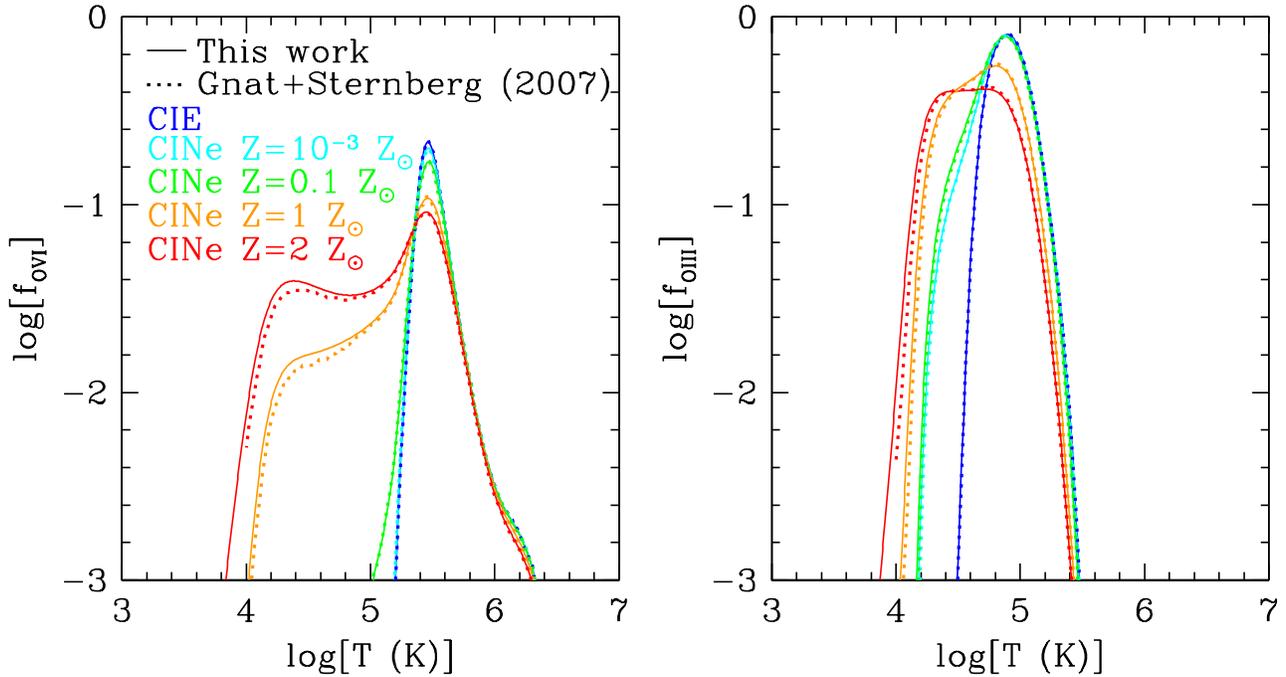}}
  \caption[]{A comparison of the predicted ion fractions of $\OVI$
    (left panel) and $\OIII$ (right panel) as a function of
    temperature and computed assuming pure collisional ionisation
    (solid lines) with those of GS07 (dotted lines, see their Figure
    3).  The gas is assumed to cool isochorically, starting out in
    ionisation equilibrium at $T=10^7$ K.  The collisional ionisation
    equilibrium case (blue) shows excellent agreement between the two
    methods, as do the non-equilibrium cases (other colours
    corresponding to different metallicities).  Even for low
    metallicity, $Z=10^{-3} \Zsolar$ (cyan), non-equilibrium effects
    of primordial species cause $\OIII$ ionisation fractions to
    diverge from CIE.  Metal abundances have been scaled to the values
    assumed by GS07.}
\label{fig:o3o6_comp}
\end{figure*}

Before considering non-equilibrium ionisation in the presence of
ionising radiation in \S\ref{sec:photo}, we will first consider only
collisional processes and compare to previous results from GS07 (who
used rates from CLOUDY ver. 06.02) and W09 (who used CLOUDY
ver. 07.02) .  Figure \ref{fig:o3o6_comp} shows $\OVI$ (left panel)
and $\OIII$ (right panel) ion fractions as a function of temperature
for cooling gas initially at $T=10^7$ K for the CIE and collisional
non-equilibrium (CINe) cases at $Z=10^{-3},0.1,1,$ and $2 \Zsolar$.
Our calculations (solid lines) are compared to GS07 (dotted lines),
who show the same two ions in their Figure 3.  We have scaled our
metal abundances to the abundances in their Table 1 for direct
comparison.  These calculations assume isochoric cooling, which is the
default unless noted otherwise.  We start in ionisation equilibrium at
$T=10^7$ K, but the results are indistinguishable if the initial gas
temperature is $5\times 10^6$ K as in GS07, because cooling times
above this temperature are long and non-equilibrium ionisation is
unimportant.

Figure \ref{fig:o3o6_comp} shows that the agreement with GS07 is good,
and demonstrates the effects of non-equilibrium processes on
ionisation fractions.  Higher metallicities result in shorter cooling
times and greater ionisation lags as recombination cannot catch up
with cooling.  Even for negligible metallicity, $10^{-3} \Zsolar$
(cyan lines), which is so low as to not impact cooling times relative
to primordial, there is non-equilibrium behaviour for $\OIII$ (right
panel).  Primordial species have non-equilibrium ionisation lags, and
GS07 demonstrated that this behaviour is independent of metallicity
below $\sim 10^{-2} \Zsolar$.  The main culprit is the ionisation lag
of recombining hydrogen.  Finally, it is worth noting that charge
transfer is important for $\OIII$.  Without recombination of $\OIII$
owing to charge exchange, $\OIII$ would persist at temperatures much
below $10^4$ K.  The slight mismatches between our method and GS07 are
likely due to the older atomic data used in CLOUDY ver. 06.02 by GS07.

\begin{figure*}
  \subfigure{\setlength{\epsfxsize}{0.98\textwidth}\epsfbox{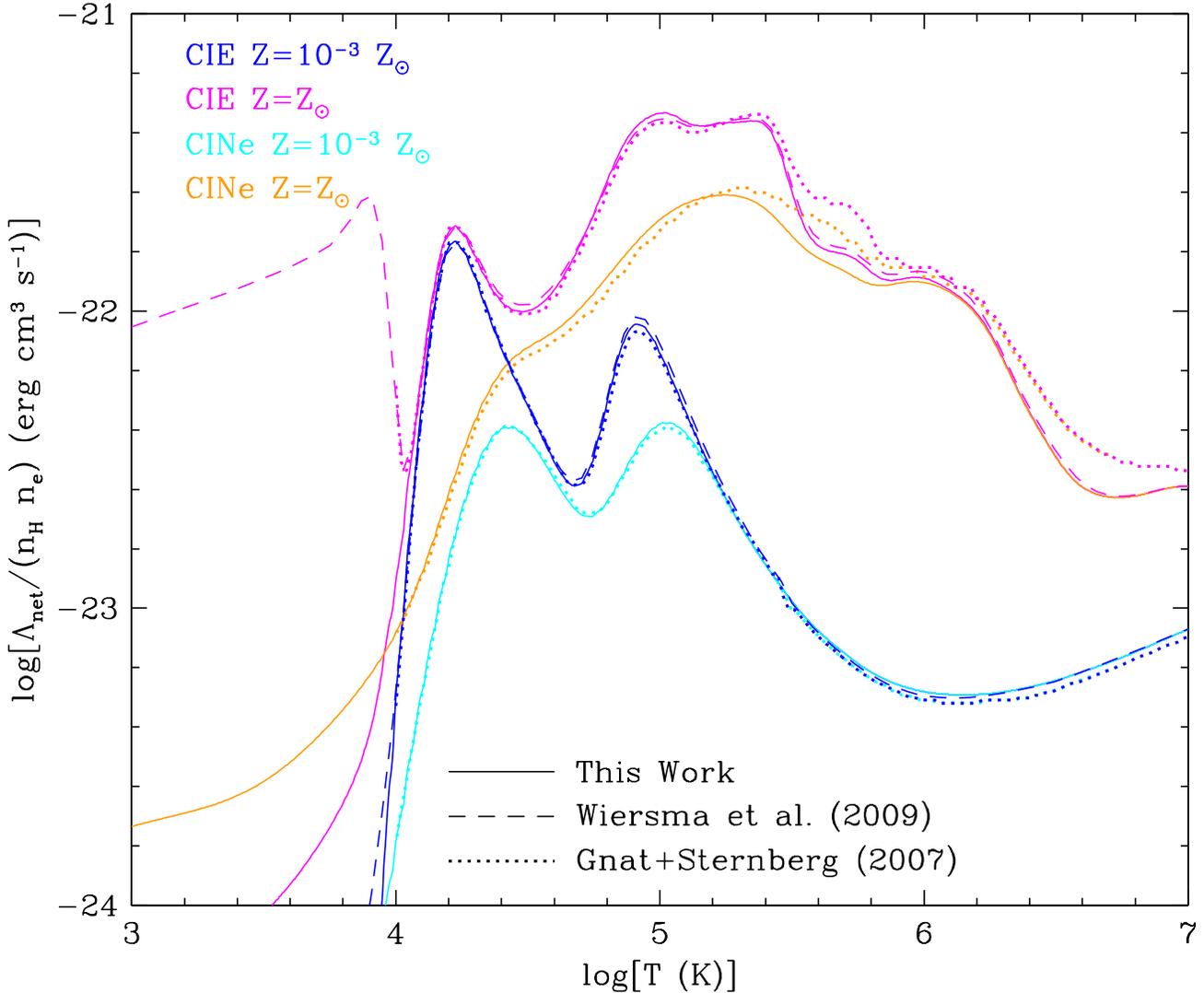}}
  \caption[]{Isochoric cooling efficiencies (plotted as
    $\Lambda_{net}/(n_{\rm H} n_e)$) for collisional ionisation
    equilibrium (CIE) and collisional ionisation non-equilibrium
    (CINe) (solid lines) compared with W09 (dashed lines) and GS07
    (dotted lines) for $10^{-3}\Zsolar$ and $\Zsolar$.  The agreement
    with W09 is excellent everywhere above $10^4$ K.  Comparison of
    our efficiencies to GS07 shows small deviations at higher
    temperatures partially owing to the different solar abundances
    used by GS07.  We find the same non-equilibrium behaviour as GS07
    of smoothed cooling curves as ion states persist down to lower
    temperatures where they generally are less efficient coolants.
    CINe curves begin in equilibrium at $T=10^7$ K, but CIE and CINe
    curves are indistinguishable at $T>10^6$ K. }
\label{fig:coolcurve_comp}
\end{figure*}

Figure \ref{fig:coolcurve_comp} shows the predicted isochoric cooling
efficiencies (solid lines), plotted as $\Lambda_{\rm net}/(n_{\rm H}
n_e)$ for CIE (blue for $10^{-3} \Zsolar$, magenta for $\Zsolar$) and
CINe (cyan for $10^{-3} \Zsolar$, orange for $\Zsolar$).  We compare
to W09 (dashed lines) for the CIE cases, and to GS07 (dotted lines) in
all cases.  We use W09 solar abundances here and throughout the rest
of the paper and note that the GS07 abundances are different.  The
differences with respect to GS07 for the cooling efficiencies at
$T\sim 10^5-10^6$ K is attributable to different neon abundances,
which are nearly twice as high for GS07.  The increased cooling at
$T>10^6$ K for GS07 is due to updated atomic data between CLOUDY
ver. 06.02 and ver. 10.00 \citep{gna12}.  These atomic updates were
included in the CLOUDY ver. 07.02 used to generate the W09 tables,
which show excellent agreement with our CIE cooling.  

In CIE, the $10^{-3} \Zsolar$ curves have peaks below $10^5$ K
corresponding to the collisional excitation of hydrogen and helium,
and a drop off at higher temperatures until Bremsstrahlung dominates
above $10^6$ K.  Solar abundances increase cooling efficiencies by
more than an order of magnitude between $T\sim 2\times 10^5$ and
$2\times 10^6$ K.  Below $T=10^4$ K, our method loses accuracy because
cooling is no longer dominated by electron-ion collisions for this
purely collisional example.  The W09 $\Zsolar$ cooling efficiency
jumps up, because ion collisions with neutral hydrogen dominate the
cooling here.  However, the cooling efficiency is much lower below
$T=10^4$ K than the W09 magenta efficiency suggests, and the jump in
the W09 rate is because we are dividing by $n_e$ (to compare with
GS07), which is very small.  Hence, the cooling rate $\Lambda_{\rm
  net}$ is very small below $10^4$ K both for W09 and us.
Nevertheless, care should be taken when our method is applied to
situations where cooling is not dominated by electron-ion collisions.

The CINe cases show the same characteristics as the corresponding GS07
efficiencies.  In non-equilibrium the high ionisation fractions for
primordial and metal species persist down to lower temperatures, where
they are generally less efficient coolants.  CINe cooling curves
therefore are both smoother and lower than the corresponding CIE
curves. This effect is also important for primordial species, which
have very peaked CIE cooling efficiencies.

Below $T=10^4$ K, our method works for CINe cases as long as
electron-ion interactions dominate the cooling, which is the case for
solar metallicity.  \citet{vas13} explore CINe cases including
molecules, finding that for $Z\ga 0.1 \Zsolar$ molecular cooling is
negligible.  Metals provide free electrons below $T<10^4$ K, which are
far more efficient coolants than molecules or ion-ion interactions.  

\begin{figure*}
  \subfigure{\setlength{\epsfxsize}{0.98\textwidth}\epsfbox{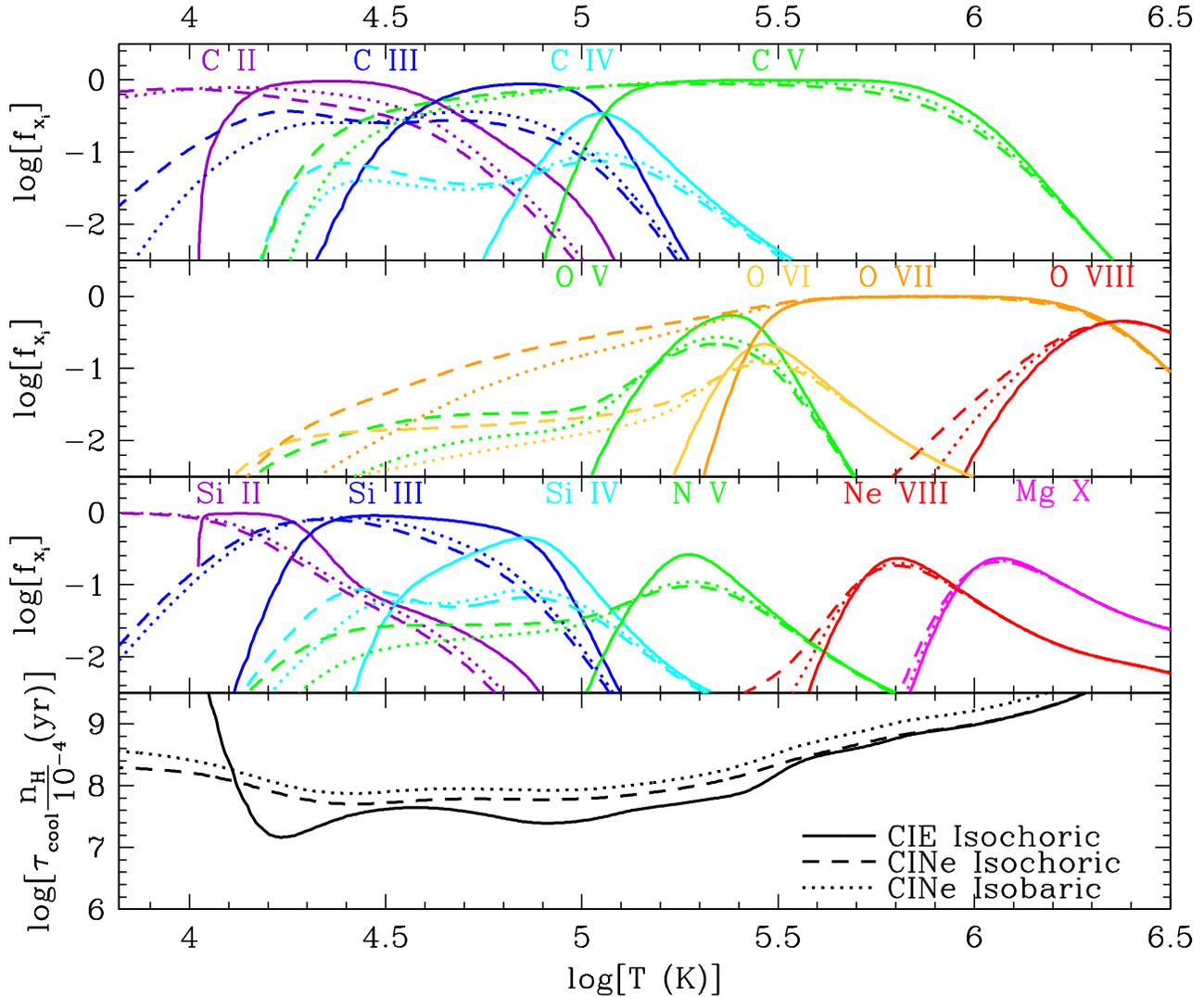}}
  \caption[]{Collisional-only ionisation fractions for the same cases
    explored in GS07: collisional ionisation equilibrium (CIE
    Isochoric, solid lines), collisional ionisation non-equilibrium at
    constant density (CINe Isochoric, dashed lines) and at constant
    pressure (CINe Isobaric, dotted lines).  We show ionisation
    fractions for some of the most common species observed in diffuse
    gas in the top three panels, as well as cooling times in the
    bottom panel to indicate how long cooling gas lives at each state,
    assuming a density of $\nh=10^{-4} \cmc$ and solar metallicity.
    The CINe Isobaric case matches the density of the CIE and CINe
    Isochoric cases at $T=10^6$ K (i.e. $\nh = 10^{-4}\cmc$($10^6$
    K)/$T$).  The CIE Isobaric case is not shown but would have
    the same ionisation fractions as the CIE Isochoric case while $\tcool$ would be a factor $5/3$ larger due to ``PdV'' work.}
\label{fig:ion_tcool_CI}
\end{figure*}

We compare the ionisation fractions for three collisional cases and
solar metallicity in Figure \ref{fig:ion_tcool_CI}: CIE Isochoric
(solid lines), CINe Isochoric (dashed lines), and CINe Isobaric
(dotted lines).  For the isobaric case, we normalise the density to
the isochoric cases at $T=10^6$ K.  A variety of observationally
obtainable metal ions are shown, starting with carbon species on top,
oxygen species in the second panel, and silicon species combined with
the Li-like ions $\NV$, $\NeVIII$, and $\MgX$ in the third panel.  Our
ion fractions compare well with GS07 (their Figure 2), exhibiting the
behaviour explained in their Figure 4.  Briefly, He-like ions ($\CV$,
$\OVII$) show the greatest ionisation lag, owing to having the longest
recombination times.  Isobaric cases do not have as great of an
ionisation lag as isochoric cases, because of their longer cooling
times at the same density due to extra ``PdV'' work.  Our CIE
ionisation fractions diverge from GS07 at $T\sim 10^4$ K (cf. $\CII$
and $\SiII$ to their Figure 2) likely owing to ion-ion collisions,
which are not included in our method, keeping the gas more ionised at
$T\la 10^4$ K.

Finally, non-equilibrium effects are most important when cooling is
most efficient, which is one reason we plot cooling times in the
bottom panel.  We use the specific case of $\nh=10^{-4} \cmc$ gas for
comparison later with the photo-ionisation case in \S\ref{sec:photo}
that depends on the density in a non-trivial manner, but these cooling
times scale inversely with $\nh$, which is why we plot $\tcool
\nh/(10^{-4}\cmc)$.  For CINe, cooling is less efficient above
$1.3\times 10^4$ but it is much more efficient at $T< 10^4$ K,
because $n_e$ is much higher owing to $\HII$ persisting down to lower
temperatures due to the recombination lag, and additionally $\HII$ is
a more efficient coolant than $\HI$ at these temperatures.  The CINe
Isobaric case has density $\nh=10^{-4}\times (10^6 {\rm K}/T) \cmc$,
which means that for the same parcel of gas, the cooling will become
faster at lower temperatures as its density rises.  However, the
cooling efficiency (erg cm$^3$ s$^{-1}$) as plotted in Figure
\ref{fig:coolcurve_comp} is lower for the CINe Isobaric case compared
to the CINe Isochoric below $T=10^4$ K owing to quicker recombination
to metal ions that are less efficient coolants.  This behaviour is also
identified by \citet{vas13} (cf. their Figure 5 for $\Zsolar$).

\begin{figure*}
  \subfigure{\setlength{\epsfxsize}{0.98\textwidth}\epsfbox{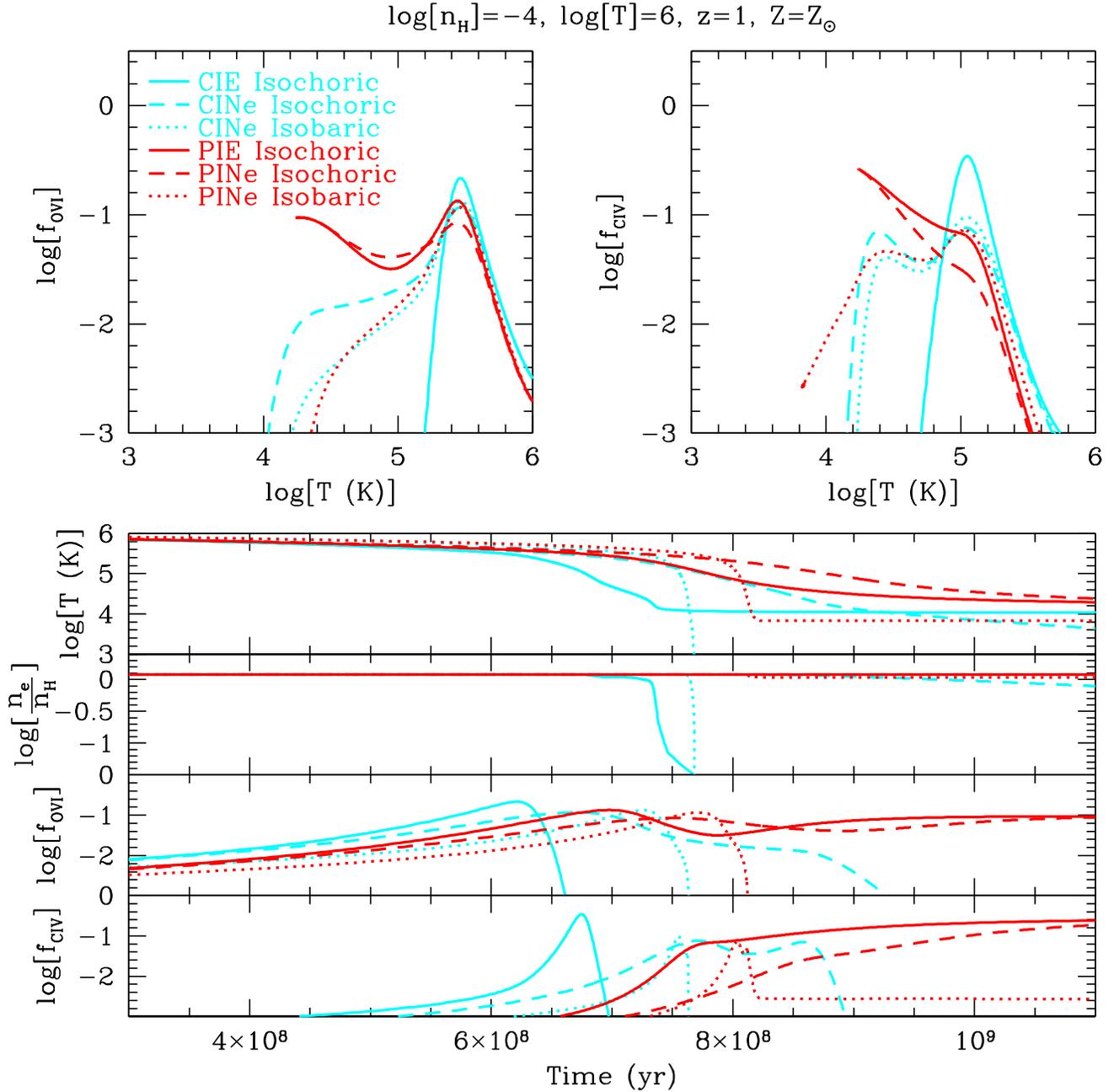}}
  \caption[]{Ion fractions as a function of temperature for $\OVI$
    (top left) and $\CIV$ (top right) for 6 cases at solar abundances:
    collisional ionisation equilibrium (CIE Isochoric, solid cyan),
    collisional non-equilibrium isochoric (CINe Isochoric, dashed
    cyan) and isobaric (CINe Isobaric, dotted cyan), photo-ionisation
    equilibrium (PIE Isochoric, solid red), photo-ionisation
    non-equilibrium isochoric (PINe Isochoric, dashed red) and
    isobaric (PINe Isobaric, dotted red).  We also show the time
    evolution from $T=10^6$ K at $t=0$ yr in the bottom four panels of
    temperature, electron density, $\OVI$ and $\CIV$ fraction.  We
    assume equilibrium initial conditions of $\nh=10^{-4} \cmc$ and
    $T=10^6$ K using the $z=1$ HM01 field in photo-ionised cases.  All
    cases achieve $T<10^5$ K temperatures in under 1 Gyr with cooling
    trajectories, electron densities, and ionisation fractions as a
    function of time differing significantly.}
\label{fig:o6c4}
\end{figure*}

\begin{figure*}
\subfigure{\setlength{\epsfxsize}{0.98\textwidth}\epsfbox{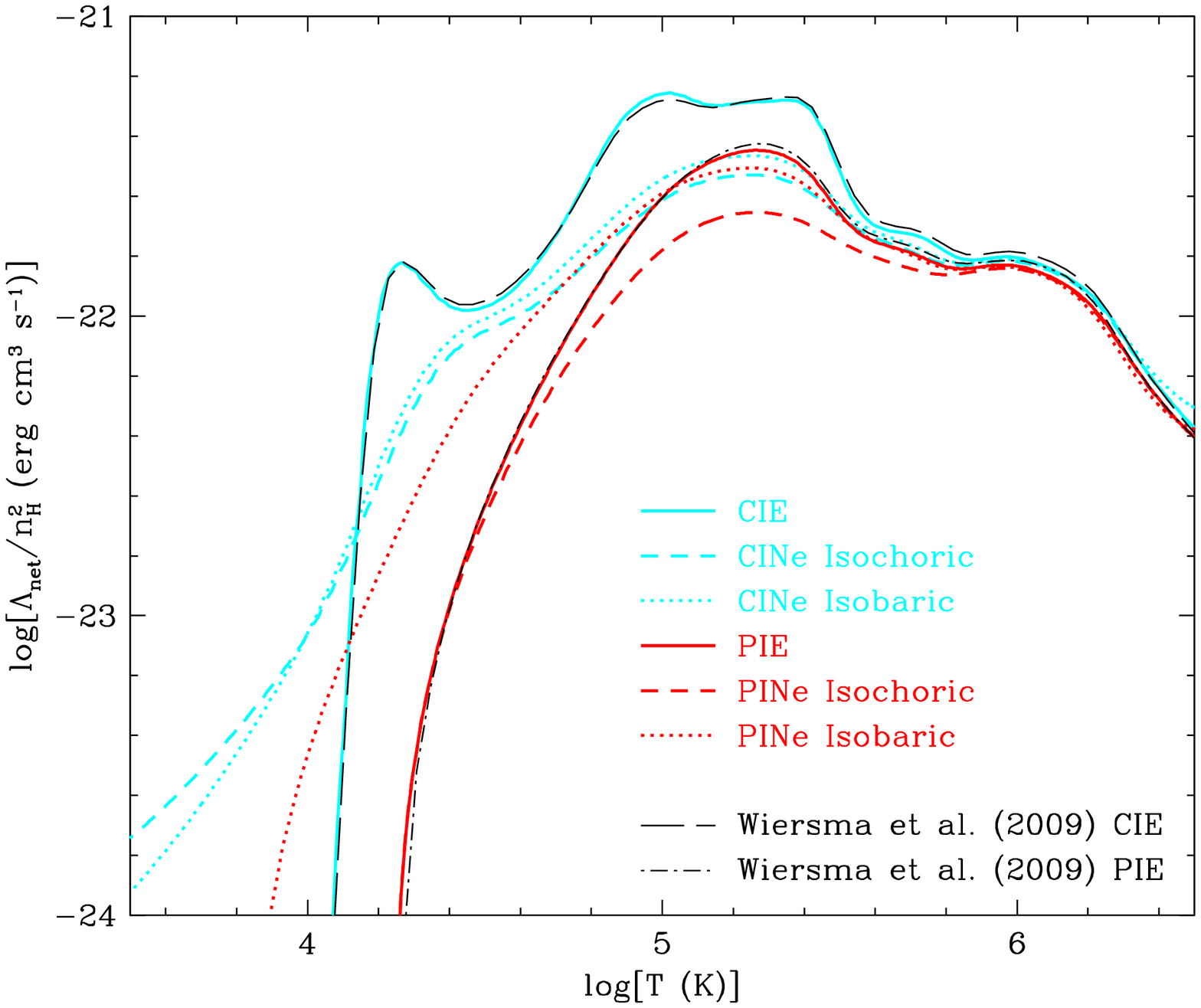}}
  \caption[]{Cooling efficiencies, plotted as $\Lambda_{\rm
      net}/\nh^2$, for the six cases displayed in Figure
    \ref{fig:o6c4} assuming solar abundances at $\nh=10^{-4} \cmc$ and
    the HM01 $z=1$ EGB.  The isobaric cases are set to $\nh=10^{-4}
    \cmc$ at $T=10^6$ K.  W09 cooling curves are displayed as thin
    dashed black lines for CIE and PIE at $z=0.957$, and agree very
    well with our equilibrium models.  Photo-ionisation suppresses
    cooling by removing bound electrons, most importantly from $\HI$.
    Non-equilibrium effects at constant density suppress cooling owing
    to recombination lags where higher ions persist to lower
    temperatures where they are less efficient coolants.  The isobaric
    cases follow the CINe Isochoric cases relatively closely, since
    the density dependence is divided out, although the PINe Isobaric
    case approaches an equilibrium temperature owing to photo-heating
    just below $T=10^4$ K and just above $\nh=10^{-2} \cmc$.
    Non-equilibrium effects on cooling are smaller in the presence of
    ionising radiation, and the effect of photo-ionisation is smaller
    for isobaric than for isochoric cooling.  }
\label{fig:coolcurve}
\end{figure*}

\section{Photo-ionised non-equilibrium metal-enriched gases} \label{sec:photo}

We now explore non-equilibrium processes in photo-ionised gases, which
is the main innovation in this paper.  Our motivation stems from the
reality that metal-enriched diffuse gases cool in the presence of an
ionising background, and that the cooling efficiencies, as well as the
observational diagnostics, are altered by the addition of
photo-ionisation to the non-equilibrium processes described in the
previous section.  The density-independence of collisional processes
is broken, because the photo-ionisation rate per unit volume due to a
uniform field scales as $n$, while collisional processes scale as
$n^2$.  Our fiducial case is a $\nh=10^{-4} \cmc$, solar abundance gas
ionised by the HM01 EGB at $z=1$.  This corresponds to an overdensity,
$\delta\sim 60$, and we will show that gas with this density cools in
less than a Hubble time.

\subsection{Non-equilibrium behaviour}

We consider the three cases explored in the previous section (CIE
Isochoric, CINe Isochoric, and CINe Isobaric) along with three new
situations with the added $z=1$ HM01 EGB: PIE Isochoric, photo-ionised
non-equilibrium (PINe) Isochoric and PINe Isobaric.  These six cases
are shown in Figure \ref{fig:o6c4}.  The gas begins in all cases in
ionisation equilibrium at $T=10^6$ K and $\nh=10^{-4} \cmc$.  The top
two panels show $\OVI$ and $\CIV$ fractions as functions of
temperature.  Below this temperature, the cooling times become short
enough to lead to non-equilibrium behaviour.

Comparing PIE to CIE, for PIE the $\OVI$ and $\CIV$ ion fractions are
much higher below $2\times10^5$ K and $6\times10^4$ K, respectively,
while these fractions are lower around the CIE peak.  All
photo-ionised curves asymptote to the equilibrium temperature where
$\Lambda_{\rm net}=0$.  Hence, if gas cools to the equilibrium point
and stabilises (e.g.\ if $\tdyn \gg \tcool$), then the gas will
maintain the ionisation fractions corresponding to the equilibrium
temperature.

Non-equilibrium effects in general allow ions to persist to lower
temperatures due to significant recombination times.  For the PINe
cases this effect is reduced as photo-ionisation also enhances the
degree of ionisation in the equilibrium case.  The impact of the
photo-ionising background is smallest for the isobaric case because
the density increases when cooling at constant pressure, thus
decreasing the ionisation parameter.

The temperature evolution in the top of the lower panels of Figure
\ref{fig:o6c4} shows the unique thermal histories of gas in the six
cases we explore.  These behaviours are also reflected in the cooling
curve plot in Figure \ref{fig:coolcurve} where we plot $\Lambda_{\rm
  net}/\nh^2$ to show net cooling efficiency for the six cases.  We
chose $\nh=10^{-4} \cmc$ at $z=1$ to show a case where
photo-ionisation has a significant effect in gas that will cool in
less than a Hubble time.  

Photo-ionisation suppresses cooling relative to the collisional
ionisation case whether in equilibrium or non-equilibrium.
Considering the isochoric cases, non-equilibrium effects retard
cooling compared to the equilibrium case resulting in the least
efficient cooling curve of the six cases (Figure \ref{fig:coolcurve})
and the hottest gas after the peak of the cooling efficiency ($T<10^5$
K) at late times (Figure \ref{fig:o6c4}).  The non-equilibrium
suppression is greater for the collisional-only cases.  Since the gas
asymptotes to a highly ionised equilibrium state in the presence of
ionising radiation, the non-equilibrium recombination lag affecting
$n_e$ and $n_{\rm HI}$ plays a smaller role and the cooling
efficiencies of PIE and PINe Isochoric become nearly the same below
$T=10^{4.5}$ K (for $\nh=10^{-4} \cmc$ at $z=1$).  Nonetheless, the
reduced peak cooling rate of the photo-ionised cases delays the
cooling of the gas by $\sim10^2$ Myr in this example.  The isobaric
non-equilibrium curves also show differences, but the effect of
photo-ionisation is smaller, because these cases achieve higher
densities at temperatures below $T=10^6$ K making photo-ionisation
less important.  The largest difference between the CINe and PINe
isobaric cases is that the latter asymptotes to an equilibrium point
after just 800 Myr when $\Lambda_{\rm net}=0$, which is achieved at a
lower temperature, $T=10^{3.8}$ K, for the isobaric density,
$\nh=10^{-1.8} \cmc$.

Figure \ref{fig:coolcurve} also shows that we reproduce the W09 PIE
cooling curve very well with our ion-by-ion cooling scheme (cf. black
thin dash-dot line and red solid line).  We re-tabulated the PIE
tables in the format of W09 using CLOUDY ver. 10.00 for both the
\citet{haa01,haa12} backgrounds, and have posted them to our website
http://noneq.strw.leidenuniv.nl.

\begin{figure}
\subfigure{\setlength{\epsfxsize}{0.49\textwidth}\epsfbox{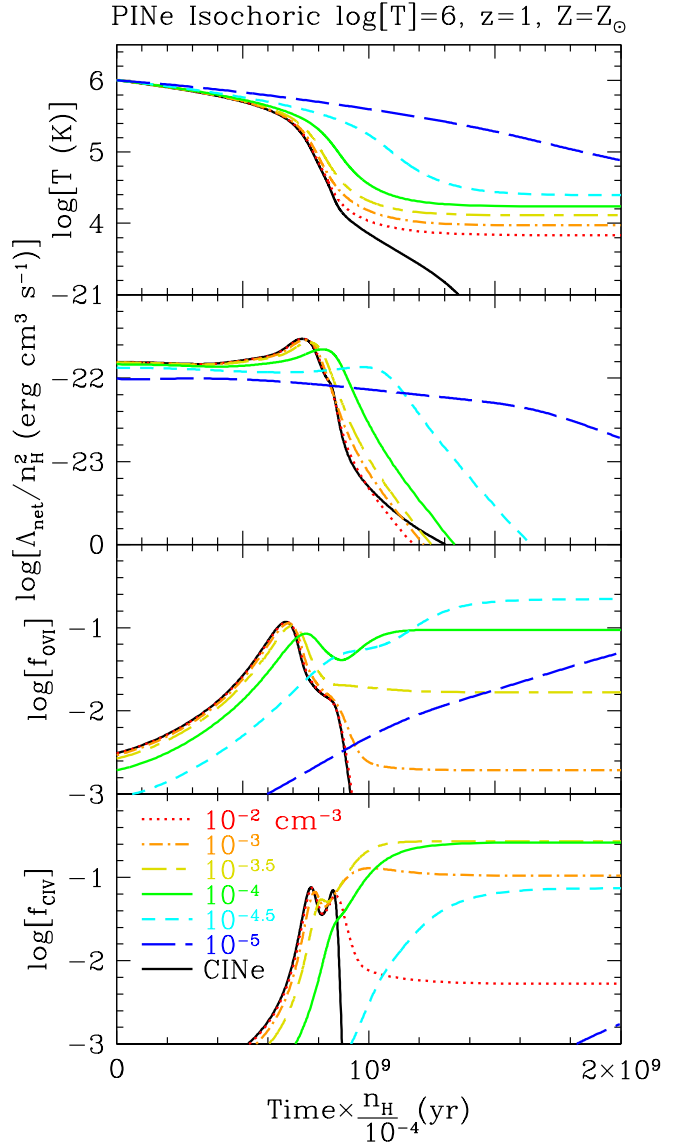}}
\caption[]{Evolution of non-equilibrium isochorically cooling solar
  abundance gas at various densities exposed to the $z=1$ HM01
  radiation field (coloured lines) and assuming collisional ionisation
  only (solid black line).  Ion fractions of $\OVI$ and $\CIV$ are
  shown along with the net cooling efficiency and temperature.  The
  time axis is scaled by $\nh$ and normalised to $\nh = 10^{-4} \cmc$
  to distill the effects of a uniform radiation field, which
  suppresses cooling more at lower densities.  The PINe cases also
  illustrate the effect of turning up or down the HM01 field strength
  at $\nh=10^{-4} \cmc$, e.g.\ the blue curve represents a $10\times$
  field increase and the red curve a $100\times$ decrease.  }
\label{fig:ion_tcool_nh}
\end{figure}

Figure \ref{fig:ion_tcool_nh} shows the dependence of the PINe
Isochoric evolution on the density.  We scaled the time axes by
$\nh^{-1}$, normalised to $10^{-4} \cmc$, to distill the effects of
photo-ionisation on the temperature and ionisation. The CINe Isochoric
case is shown in black for comparison.  The effect of photo-ionisation
is smallest at the highest densities.  Lowering the density results in
more delayed cooling even after removing the $\nh^{-1}$ dependence of
the cooling time, as well as higher equilibrium temperatures
(i.e.\ $\Lambda_{\rm net} = 0$) because the photo-heating rate
decreases less rapidly with the density than the cooling rate in this
regime.  The $\nh = 10^{-5} \cmc$ case has the most extreme
suppression of cooling shown, yielding a cooling time from $10^6$ to
$10^5$ K of 19 Gyr, i.e. longer than a Hubble time.  The EGB would
therefore evolve, becoming weaker below $z=1$, but we have turned off
this evolution for simplicity.  Finally, we note that Figure
\ref{fig:ion_tcool_nh} also shows the effect of adjusting the strength
of the HM01 field by a factor of $10^{-4} \cmc/\nh$ at fixed density
if we keep the time axis fixed in absolute terms.  For example, the
blue $\nh=10^{-5} \cmc$ curve is applicable to a $\nh=10^{-4} \cmc$
gas cooling in a $10\times$ stronger field and down to $10^5$ K in 1.9
Gyr.

\begin{figure}
\subfigure{\setlength{\epsfxsize}{0.45\textwidth}\epsfbox{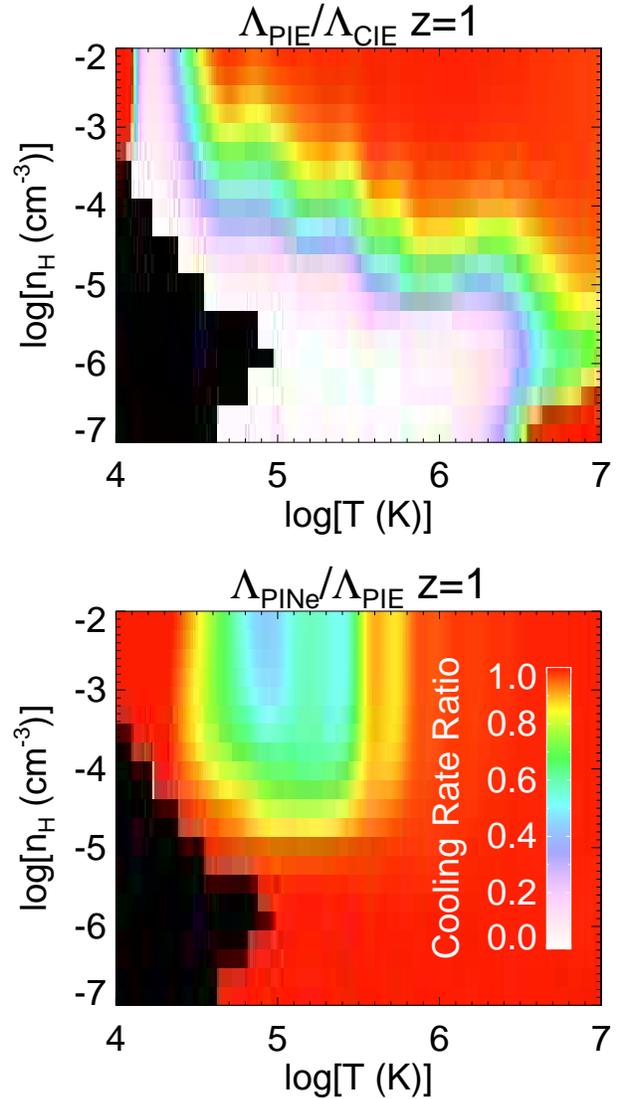}}
\caption[]{The ratio of net cooling rates as a function of temperature
  and density for $Z=\Zsolar$ at $z=1$ assuming isochoric conditions
  beginning at $T=10^7$ K.  The top panel shows the ratio of PIE to
  CIE cooling rates, indicating that photo-ionisation suppresses the
  cooling rate more at lower density (higher ionisation parameter).
  The bottom panel shows the ratio of PINe to PIE cooling rates,
  indicating that non-equilibrium processes affect cooling most at
  higher densities between $T\sim 10^{4.5}-10^{6}$ K.  Cooling at high
  temperatures and low densities is dominated by Compton cooling,
  which has been subtracted here, but makes the comparison in the
  bottom right corner of the top panel inaccurate.  Black corresponds
  to conditions where there is net heating.}
\label{fig:tcoolratio_density}
\end{figure}

The ratios of isochoric cooling rates are displayed in Figure
\ref{fig:tcoolratio_density} as a function of density and temperature
assuming solar enriched gas irradiated by the $z=1$ HM01 EGB,
initially at $T=10^7$ K.  The relative cooling rates of PIE cooling
versus CIE appear in the top panel, indicating that photo-ionisation
suppresses the cooling rate more as the density goes down (or the
ionisation parameter increases), as is demonstrated by W09.  The
bottom panel shows the ratio of PINe and PIE cooling rates to show
where non-equilibrium effects become important for gas cooling
dynamics.  Non-equilibrium effects are more important at higher
densities corresponding to circumgalactic halo gas, where
photo-ionisation is weaker.  At even higher densities than plotted
here, the PINe/PIE cooling rate ratios reflect the CINe versus CIE
case, as shown by \citet{vas11}.

To summarise, photo-ionisation reduces cooling efficiencies and
increases the degree of ionisation for both equilibrium and
non-equilibrium cooling.  Non-equilibrium effects on the evolution of
the cooling rate and on the ionisation balance are smaller in the
presence of ionising radiation.  The effect of photo-ionisation is
smaller for isobaric than for isochoric cooling, which also agrees
with the conclusions of \citet{vas11}.

\subsection{Metal-line diagnostics}

QAL metal-line observations are used to diagnose both the physical
state of the gas (e.g. its temperature, ionisation parameter, and
chemical composition) and the dynamical state of the gas
(e.g.\ radiative cooling, conductive interface, turbulent mixing
layer, shock front).  Calculations usually assume equilibrium
ionisation or ignore the effect of photo-ionisation, which is why we
emphasise the need to include both non-equilibrium and
photo-ionisation in diagnosing metal lines.

The bottom two panels of Figure \ref{fig:o6c4} demonstrate the
evolution of the $\OVI$ and $\CIV$ ion fractions for our six fiducial
cases and the bottom panels of Figure \ref{fig:ion_tcool_nh} show the
evolution for different densities assuming PINe Isochoric cooling.
Note that for radiatively cooling gas, the probability of observing
the ion ratios at the peak of the cooling curve is small compared to
the probability of observing the gas at either higher temperatures
where cooling is slower, or at lower temperatures when the gas
approaches thermal equilibrium when $\Lambda_{\rm net}=0$.

\begin{figure*}
  \subfigure{\setlength{\epsfxsize}{0.98\textwidth}\epsfbox{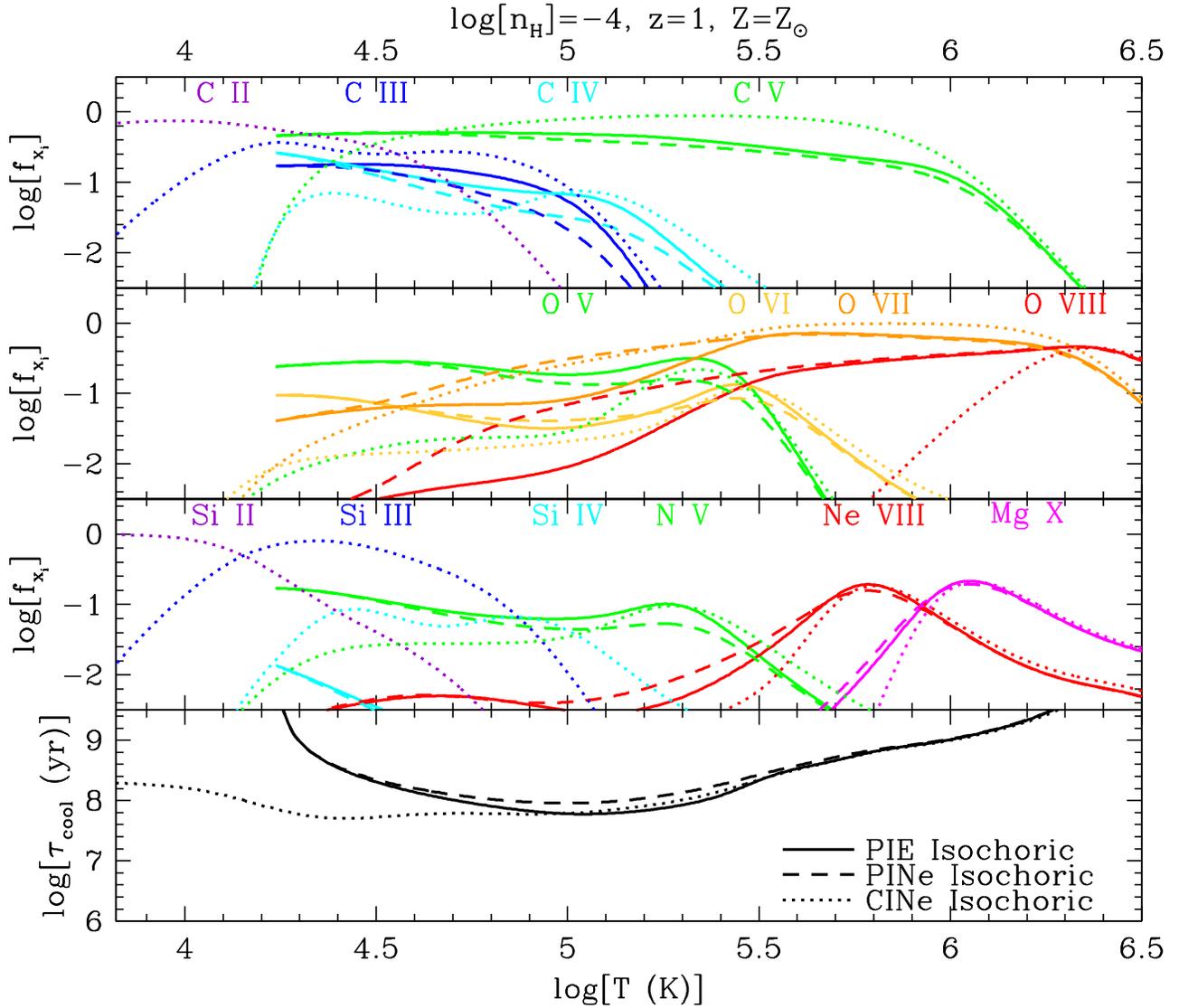}}
\caption[]{Ionisation fractions as a function of temperature for some
  of the most common species observed in isochoric, radiatively
  cooling diffuse gas along with cooling times (bottom panel) to
  indicate how long cooling gas lives at each state assuming
  $\nh=10^{-4} \cmc$.  We assume an initial temperature of
  $T=10^{6.5}$ K.  The density is fixed at $\nh=10^{-4} \cmc$ at all
  times.  We assume solar abundances and, for the photo-ionised case
  (solid and dashed curves) the presence of the $z=1$ HM01 EGB.  In
  contrast with the analogous plot for collisional-only ionisation
  (Figure \ref{fig:ion_tcool_CI} and dotted curves in this Figure), a
  given ionisation species will often persist to lower temperatures
  due to photo-ionisation and stop at $T=10^{4.1}$ K owing to
  photo-heating balancing cooling.  For isobaric cooling (not shown
  here) the differences between PINe and CINe are smaller than for
  isochoric cooling.}
\label{fig:ion_tcool_PI}
\end{figure*}

In Figure \ref{fig:ion_tcool_PI}, we plot ion fractions for a variety
of observed UV resonance transitions observed in QAL spectra for the
PIE and PINe Isochoric cases for gas with $\nh=10^{-4} \cmc$ that is
exposed to the $z=1$ EGB, as well as the CINe Isochoric case.
Allowing for the presence of ionising radiation opens up a new set of
solutions with this figure showing only one density for a specific
EGB.  Entirely new families of solutions are possible for varying
densities and backgrounds, which is why we offer the ionisation
fraction tables for several densities for the HM01 background at
several redshifts on our website http://noneq.strw.leidenuniv.nl, and
include the solutions also for the newer \citet{haa12} background.
Additionally, we provide cooling efficiencies, $\Lambda_{\rm
  net}/\nh^2$, and the $\tcool$ as a function of temperature for
various densities for the PIE Isochoric, PINe Isochoric, and PINe
Isobaric cases.

Figures \ref{fig:o6c4}, \ref{fig:ion_tcool_nh}, and
\ref{fig:ion_tcool_PI} show that assuming CINe rather than PINe can
easily lead to order of magnitude errors in the predicted ion
fractions.  An example of an application for which the difference is
much smaller is a QAL component absorber with $N_{\rm NeVIII} \sim
N_{\rm NV}$.  This is very close to what has been observed by
\citet{tri11} who constrain the temperature of a radiatively cooling
gas based primarily on these two species having similar columns
(within 0.5 dex of each other) in various components of a single
system.  Since the solar number density abundances of nitrogen and
neon are similar, $N_{\rm NeVIII} \sim N_{\rm NV}$ means that $f_{\rm
  NeVIII} \sim f_{\rm NV}$.  The rapid change in the $\NV$ and
$\NeVIII$ fractions as a function of temperature visible in the third
panel of Figure \ref{fig:ion_tcool_PI} provides a powerful constraint
for a single-phase radiatively cooling gas model of $T\sim 10^{5.5}$
K.  By seeing where the respective green and red lines cross in this
panel, we can see that photo-ionised models predict only slightly
lower temperatures ($T=10^{5.46}$ K for PINe Isochoric) than for the
collisional-only case ($T=10^{5.58}$ K).  The ionisation correction
also decreases for the PINe case as the ionisation fractions are $0.2$
dex higher than for CINe meaning that taking photo-ionisation into
account yields a lower gas mass.

However, we provide a cautionary note about applying a single-phase
radiatively cooling solution to this situation, because the
probability of catching gas cooling at a temperature where $N_{\rm
  NeVIII} \sim N_{\rm NV}$ is low owing to $\tcool$ being short
(bottom panel of Figure \ref{fig:ion_tcool_PI}).  Gas at these
temperatures can exist, but it is a priori much more probable to
observe cooling gas at either lower temperatures where $\tcool$
approaches infinity at the equilibrium temperature, or at $T\ga 10^6$
K where $\tcool$ is longer.  A multi-phase model comprising of
radiatively cooling sub-clumps occupying a temperature distribution
weighted by $\tcool(T)$ between $T\sim 10^4 - \ga 10^6$ K is more
conceivable.  However, this model would fail likely owing to the
hotter sub-clumps having significant $\MgX$ absorption violating the
upper limit constraint observed by \citet{tri11}.  A related
possibility is that metal-enriched gas has a maximum of $T\sim 10^{5.5}$ K
accompanied by a much larger reservoir at lower temperatures, which
would not violate the $\MgX$ upper limit.  However, we provide an
alternative model in Oppenheimer \& Schaye (2013), which explains the
observations by appealing to enhanced fossil photo-ionisation from a
local AGN.

\subsection{Ionisation background variations}

\begin{figure}
  \subfigure{\setlength{\epsfxsize}{0.45\textwidth}\epsfbox{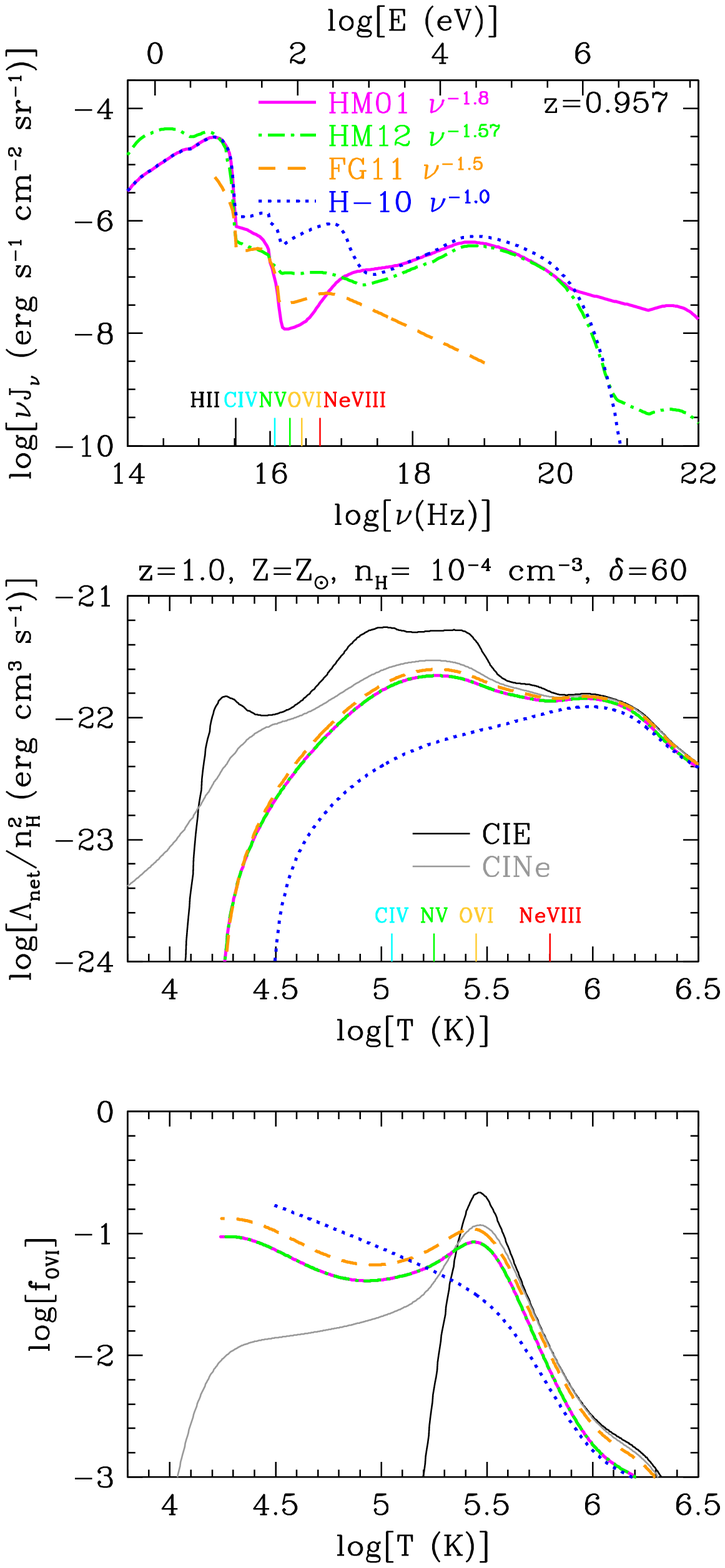}}
\caption[]{Demonstration of how different ionising backgrounds (top
  panel) affect radiative cooling efficiencies (middle panel) and ion
  fractions ($\OVI$ shown in bottom panel).  Different versions of the
  EGB at $z=0.957$ show how changing the EUV slope can dramatically
  alter the ionising background at frequencies of commonly observed
  Lithium-like ion potentials.  Non-equilibrium isochoric cooling
  curves for solar metallicity gas with density $\nh=10^{-4} \cmc$
  beginning at $T=10^{6.5}$ K at $z=1$ (plotted as $\Lambda_{\rm
    net}/\nh^2$) show that harder radiation fields suppress cooling
  more by ionising metal coolants to a higher degree.  For all
  backgrounds the cooling efficiency and $\OVI$ fraction are very
  different from the CIE and CINe cases (also shown in the bottom two
  panels).  }
\label{fig:bkgd_cool_fovi}
\end{figure}

Cooling efficiencies and QAL diagnostics both depend on the spectral
shape and strength of the EGB.  Thus far, we have used the HM01 EGB
because it is widely used \citep[e.g.][]{sch10,opp12a} and because it
fits observed metal ion ratios at $z\sim 3$ \citep{agu08}, as well as
the full $\HI$ column density distribution \citep{dav10, alt11}.
However, the intensity and especially the shape of the EGB are rather
uncertain as they depend on poorly constrained escape fractions,
intrinsic spectral slopes, and extrapolations of luminosity functions.
Recently published backgrounds include the \citet[][hereafter
  HM12]{haa12} and the \citet[][hereafter FG11\footnote{Using the
    Dec. 2011 update found at
    https://www.cfa.harvard.edu/$\sim$cgiguere/UVB.html}]{fau09}
model.  We plot these backgrounds at the output nearest to $z=1$ in
the top panel of Figure \ref{fig:bkgd_cool_fovi} to demonstrate the
variations in the EGB, especially at the extreme UV (EUV) energies
that correspond to the ionisation potentials of commonly observed
Lithium-like ions indicated in the Figure.  The assumed QSO EUV slope
($>$13.6 eV) ranges from $\nu^{-1.5}$ to $\nu^{-1.8}$ in the above
listed models and can change the strength of the field by more than an
order of magnitude at the ionisation energy of $\OVI$ (114 eV).  Given
that the observed QSO EUV slope ranges between at least $\nu^{-3.0}$
and $\nu^{0.0}$ \citep{tel02, shu12}, we generate an EGB with a harder
input QSO EUV slope, $\nu^{-1.0}$, using a post-2001, pre-2012 version
of the CUBA package (F. Haardt, private communication).  This
background (H-1.0) is shown as the dotted blue line in Figure
\ref{fig:bkgd_cool_fovi}.  This EGB may be relevant for diffuse metals
in regions occupied by harder than normal QSOs.

Considering our fiducial case ($\nh=10^{-4} \cmc$ at $z=1$ and solar
abundances), we show the PINe Isochoric cooling efficiencies for these
four backgrounds in the middle panel of Figure
\ref{fig:bkgd_cool_fovi}.  The cooling time from $10^6\rightarrow
10^5$ K, i.e.\ near the peak of the cooling curve, is 675 Myr for CIE
and 773 Myr in CINe.  The addition of photo-ionisation from the three
published EGBs increases this cooling time to between 820 and 903 Myr.
The more extreme H-1.0 EGB heavily suppresses metal coolants at these
temperatures, increasing $\tcool$ to 1.47 Gyr.  Hence, the inclusion
of non-equilibrium effects and photo-ionisation from a hard EGB can
double the cooling timescale.  Since the accretion of gas depends on
its ability to cool, the process of galaxy formation can be
significantly altered by photo-ionisation.  This conclusion will only
be strengthened if local sources of ionising radiation increase the
intensity of the radiation field above that of the EGB.  Such effects
have been argued to be important \citep[e.g.][]{sch06, can10, rah13},
but we have conservatively ignored them here.

While the most extreme EGB reduces cooling efficiencies by a factor of
two, the diagnostics of observed QAL can change much more as
illustrated by the $\OVI$ fractions in the bottom panel of Figure
\ref{fig:bkgd_cool_fovi}.  Below the peak of the CIE cooling curve
(i.e.\ at $T<10^{5.3}$ K), $\OVI$ fractions are much higher for PINe
than for either CIE or CINe.  The difference between the different
PINe cases is smaller ($\approx 2\times$).  We emphasise once again
that, a priori, radiatively cooling gas is most likely to be observed
when $\tcool$ is longest, i.e.\ at the temperatures near the
equilibrium point ($T\sim 20,000-30,000$ K).

In a cooling rate ratio plot analogous to Figure
\ref{fig:tcoolratio_density}, we vary the redshift of the HM01 EGB,
while keeping $\nh=10^{-4.0} \cmc$ and assuming $Z=\Zsolar$.  The top
panel shows that PIE cooling is suppressed more than the CIE case at
higher redshift, with the difference becoming less from
$z\sim2\rightarrow 0$ as the EGB weakens.  The ratio of PINe to PIE
cooling in the bottom panel shows that non-equilibrium cooling
suppression is greatest at $z=0$ and $z=6$, and reaches a minimum at
$z\sim 2.5$ when the background is strongest.  A similar plot was made
by \citet{vas11} (cf. their Figure 9), showing the same trend, but at
a lower density ($n=10^{-4.0} \cmc$ or $\nh\sim 10^{-4.3} \cmc$).  In
general, as the ionisation parameter increases, the cooling
suppression owing to photo-ionisation becomes more important than the
suppression owing to non-equilibrium effects.

\begin{figure}
\subfigure{\setlength{\epsfxsize}{0.45\textwidth}\epsfbox{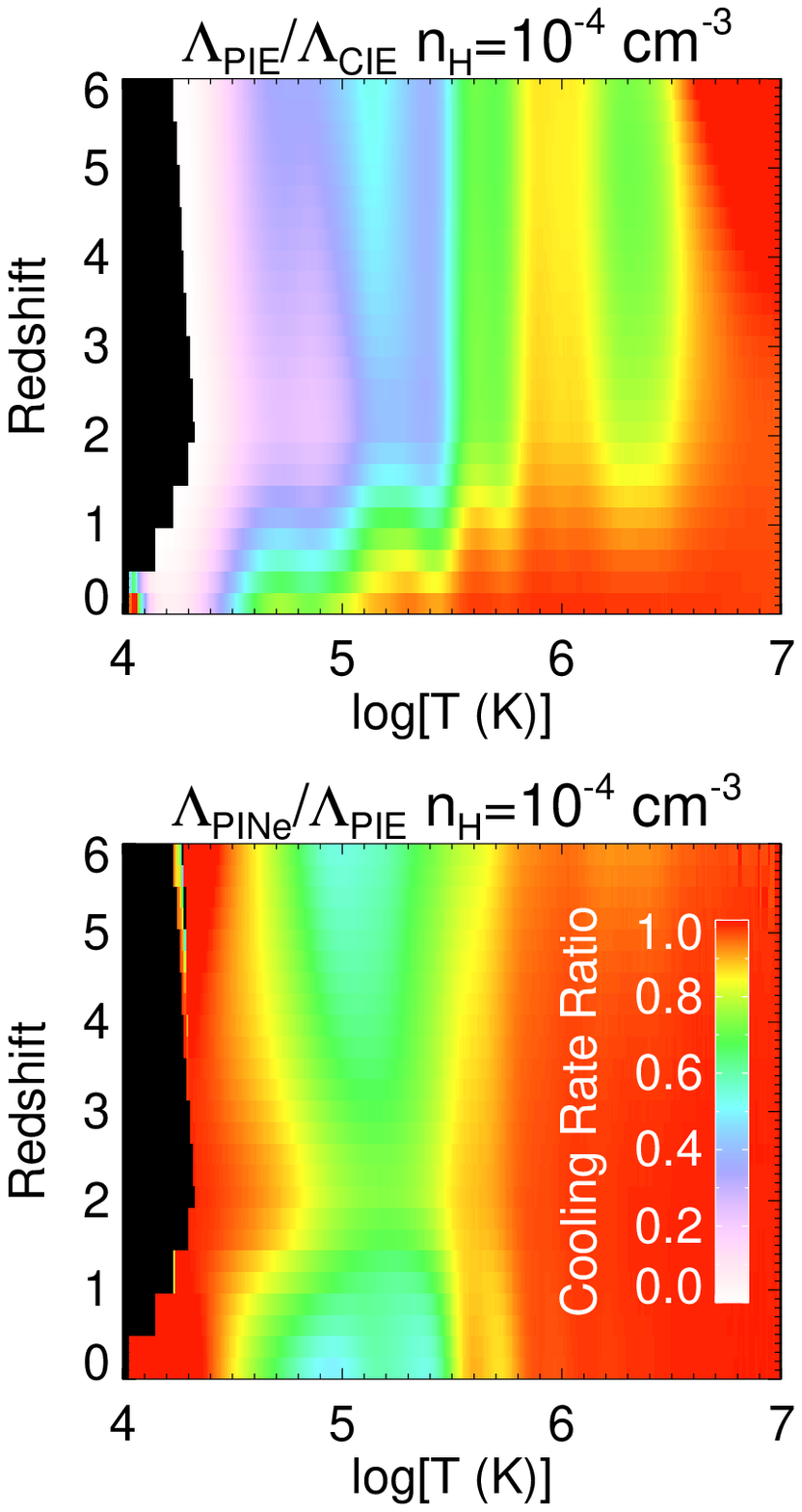}}
\caption[]{The ratio of net cooling rates as a function of temperature
  and redshift for $Z=\Zsolar$ at $\nh=10^{-4.0} \cmc$ assuming the
  HM01 EGB.  The top panel shows the ratio of PIE to CIE cooling
  rates, indicating that photo-ionisation suppresses more at higher
  redshift when the EGB is stronger.  The bottom panel shows the ratio
  of PINe to PIE cooling rates, indicating that non-equilibrium
  processes affect cooling between $T\sim 10^{4.5}-10^{6}$ K at all
  redshifts, and the most when the EGB is weaker.  Black corresponds
  to conditions where there is net heating. }
\label{fig:tcoolratio_redshift}
\end{figure}

\section{Summary}

Radiative cooling is an essential ingredient for any model of the
formation of galaxies and the evolution of the intergalactic
medium. Cooling efficiencies depend on the chemical composition and
the ionisation balance. Knowledge of the ionisation balance is also
critical for the interpretation of observations of quasar absorption
lines.

Although simulations nearly always assume ionisation equilibrium, this
assumption will generally break down in gas that is cooling rapidly or
that is exposed to a fluctuating radiation field. In particular,
recombination lags may leave the gas over-ionised, thus reducing the
cooling efficiency \citep[e.g.][]{kaf73,gna07}.

Diffuse gas is exposed to the extra-galactic background radiation and
possibly also to radiation from local sources. The presence of UV and
X-ray radiation can strongly boost the ionisation of the gas and must
therefore be taken into account when interpreting observations. The
resulting removal of bound electrons injects energy
(i.e.\ photo-heating), but also suppresses the cooling efficiencies of
gas hot enough to cool through collisional excitation
\citep[][i.e.\ $T > 10^4$ K]{efs92,wie09a}. Simulations have only
recently begun to take these effects into account for metals
\citep[e.g.][]{sch10,for13}, although still under the assumption of
ionisation equilibrium.

Clearly, a realistic treatment of ionisation and cooling can neither
assume ionisation equilibrium nor ignore the effects of the ionising
background. In addition, the effects of non-equilibrium and
photo-ionisation must be considered for both primordial and heavy
elements.

We have presented a method for calculating non-equilibrium ionisation
and cooling of diffuse gas exposed to ionising radiation.  We consider
133 ionisation states of the 11 species that dominate the cooling
efficiencies of diffuse gas (H, He, C, N, O, Ne, Mg, Si, S, Ca, \&
Fe). Cooling and photo-heating efficiencies are computed ion-by-ion using
tables generated with CLOUDY. After each time step, the ionisation
balance is updated using a reaction network that includes radiative
and di-electric recombination, collisional ionisation,
photo-ionisation, Auger ionisation, and charge transfer. We make
available the equilibrium and non-equilibrium calculations presented
here in table format at http://noneq.strw.leidenuniv.nl. We also
provide ionisation, recombination, cooling, and photo-heating
coefficient input tables that can be integrated into hydrodynamic
simulations, and explain our method to use these in such calculations.

We demonstrated that our method reproduces published collisional
ionisation equilibrium and non-equilibrium ionisation fractions and
cooling efficiencies \citep{gna07}, and photo-ionised equilibrium
ionisation fractions \citep{fer98} and cooling efficiencies
\citep{wie09a}.  We also reproduce the basic results of \citet{vas11},
who also calculated non-equilibrium ionisation and cooling in the
presence of an extra-galactic background (EGB).

Like \citet{gna07}, we found that non-equilibrium cooling is confined
to $T<10^{6.7}$ K.  At these temperatures, the recombination time
becomes significant compared to the cooling time, resulting in a
recombination lag and reduced cooling efficiencies. 

Photo-ionisation and non-equilibrium effects both tend to boost the
degree of ionisation and to reduce cooling efficiencies, sometimes by
significant factors. The effect of the EGB is larger for lower
densities (i.e.\ higher ionisation parameters). Hence,
photo-ionisation affects (equilibrium and non-equilibrium) cooling
more under isochoric than under isobaric conditions. Non-equilibrium
effects are smaller in the presence of an EGB and may therefore have
been overestimated by previous work.

Observational diagnostics of diffuse, metal-enriched gases
(e.g.\ metal absorption lines probed in quasar sight lines) are
altered even more significantly than the cooling efficiencies, both
under equilibrium and non-equilibrium conditions. Hence, the inclusion
of the EGB opens up an entire new range of ionisation solutions
depending on gas density, metallicity, and the strength and shape of
the ionising radiation field.

We emphasised that not all ionisation solutions of a cooling gas
should be treated with equal probability.  For example, applying an
ionisation solution to an observed absorption system that corresponds
to a temperature near the peak of the cooling curve (i.e.\ where
cooling times are shortest) assumes all the gas traced by the absorber
has the solution that is a priori least probable.  The addition of an
EGB photo-heats the gas, usually leading to an equilibrium temperature
where net cooling, radiative cooling minus photo-heating, is zero.
Since the cooling time is effectively infinite at this point, this
equilibrium solution with $T\sim 10^4$ K, has a high probability.

Considering the dynamics of cooling is critical when applying a
time-dependent solution to metal absorbers, which often require
multi-phase models consisting of multiple parcels of gas at different
temperatures. Hence, we provide dynamical cooling times and
efficiencies with all the ionisation fractions included on our
website. Our fiducial case of solar abundance gas at $\nh=10^{-4}
\cmc$ irradiated by a $z=1$ EGB predicts much higher $\CIV$ and $\OVI$
fractions than pure collisional ionisation models.

We include coefficient lookup tables extending from $T=10^{2}-10^{9}$
K.  Our method is applicable to cases when electron-ion collisions
dominate the cooling, which includes nearly all intergalactic and
circumgalactic gas.  We demonstrate that our method works for
metal-enriched gas at $T<10^4$ K that is ionised by the EGB.  However
our method is not accurate for $T\leq 10^4$ K if $\HI$ is nearly fully
neutral, e.g.\ in the dense parts of the interstellar medium, because
ion-ion interactions, molecules and dust may then be important.

We encourage the use of our website which provides photo-ionised,
non-equilibrium ionisation and cooling data as a function of
temperature for a larger range of redshifts, metallicities, and
densities, including both the \citet{haa01} and (2012) backgrounds,
than presented here.  We will continue to update this website with the
release of new versions of CLOUDY containing updated atomic data
\citep[e.g.][]{lyk13}.  We also provide collisional and equilibrium
results and provide some examples of how the calculations can be
compared to observations.  

Diagnosing the physical state of the metal-enriched intergalactic and
circumgalactic media where cooling times are short requires tracking
non-equilibrium ionisation in the presence of the ubiquitous
extra-galactic background.

\section*{Acknowledgements}

We are grateful to Orly Gnat, Alex Richings, Chris Churchill, Kristian
Finlator, Mike Shull, and Evgenii Vasiliev for stimulating discussions
and assistance.  We acknowledge Gary Ferland and the CLOUDY community
for providing an essential tool in this work, and Francesco Haardt for
providing the CUBA package to generate unique ionisation backgrounds.
This work benefited from financial support from the Netherlands
Organisation for Scientific Research (NWO) through VENI and VIDI
grants, from NOVA, from the European Research Council under the
European Unions Seventh Framework Programme (FP7/2007-2013) / ERC
Grant agreement 278594-GasAroundGalaxies and from the Marie Curie
Training Network CosmoComp (PITN-GA-2009-238356).

\appendix

\section{Equilibrium ionisation comparisons with CLOUDY} \label{sec:ioncompare}

\begin{figure*}
  \subfigure{\setlength{\epsfxsize}{0.32\textwidth}\epsfbox{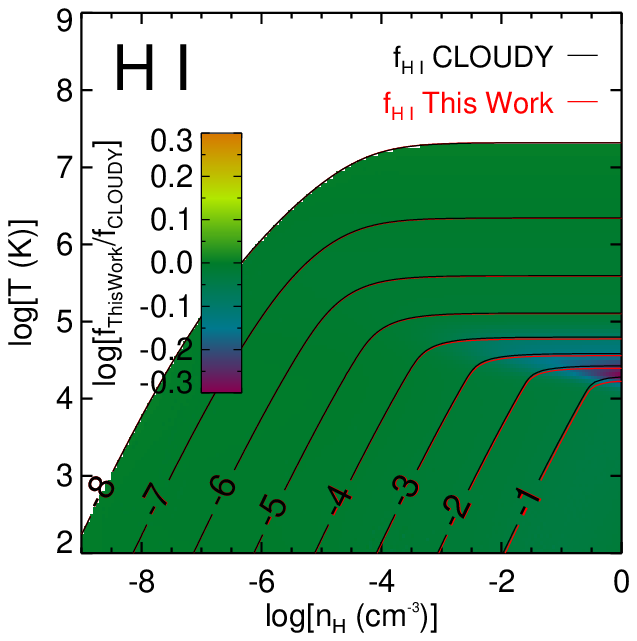}}
  \subfigure{\setlength{\epsfxsize}{0.32\textwidth}\epsfbox{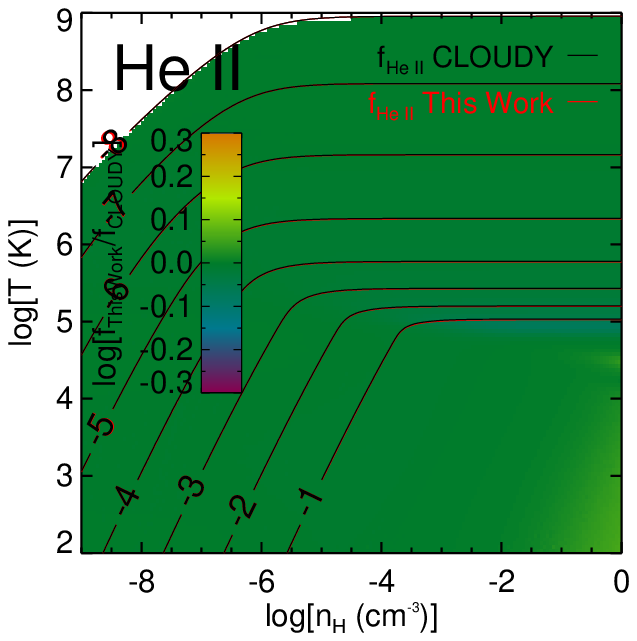}}
  \subfigure{\setlength{\epsfxsize}{0.32\textwidth}\epsfbox{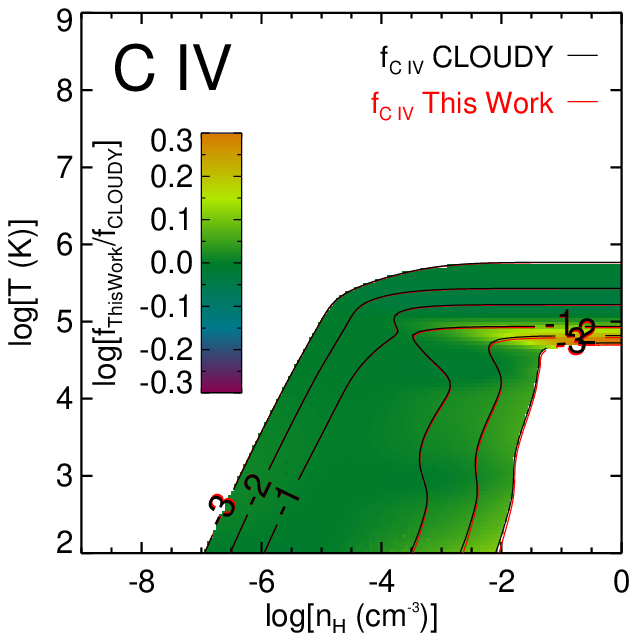}}
  \subfigure{\setlength{\epsfxsize}{0.32\textwidth}\epsfbox{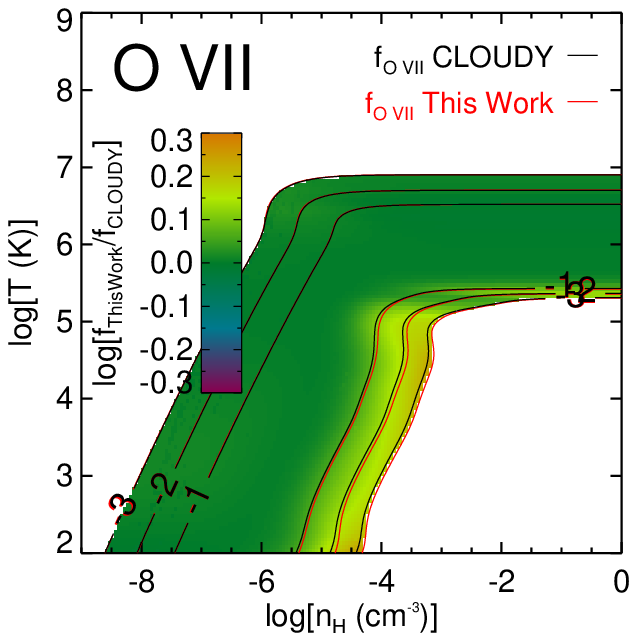}}
  \subfigure{\setlength{\epsfxsize}{0.32\textwidth}\epsfbox{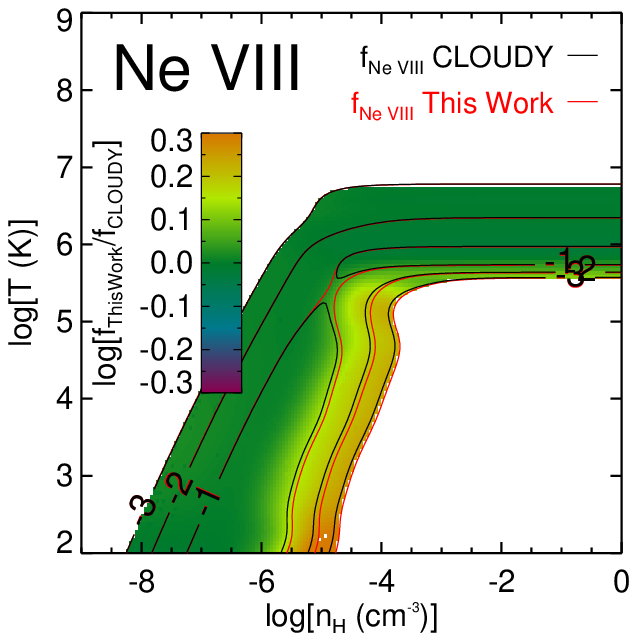}}
  \subfigure{\setlength{\epsfxsize}{0.32\textwidth}\epsfbox{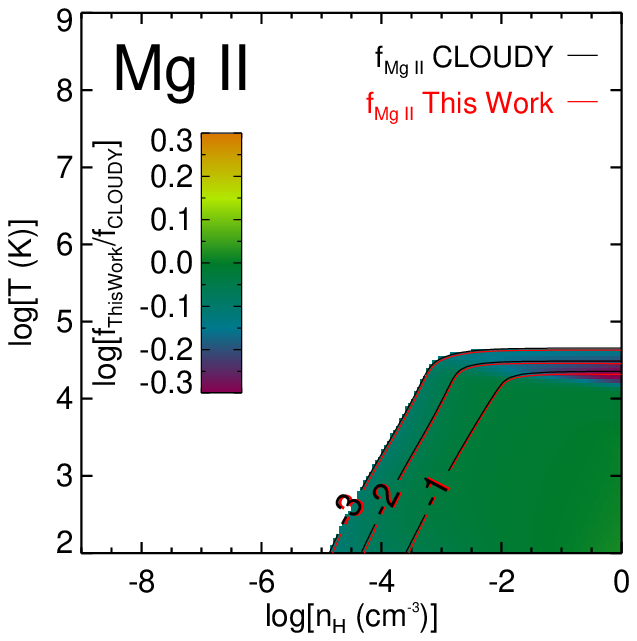}}
  \subfigure{\setlength{\epsfxsize}{0.32\textwidth}\epsfbox{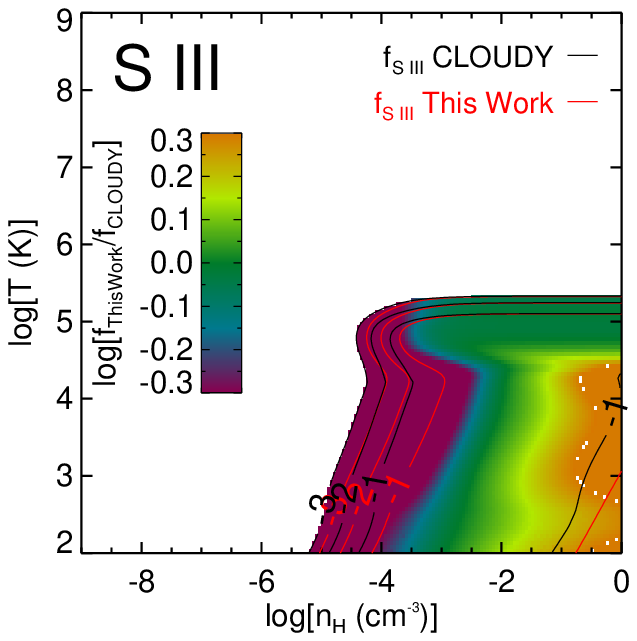}}
  \subfigure{\setlength{\epsfxsize}{0.32\textwidth}\epsfbox{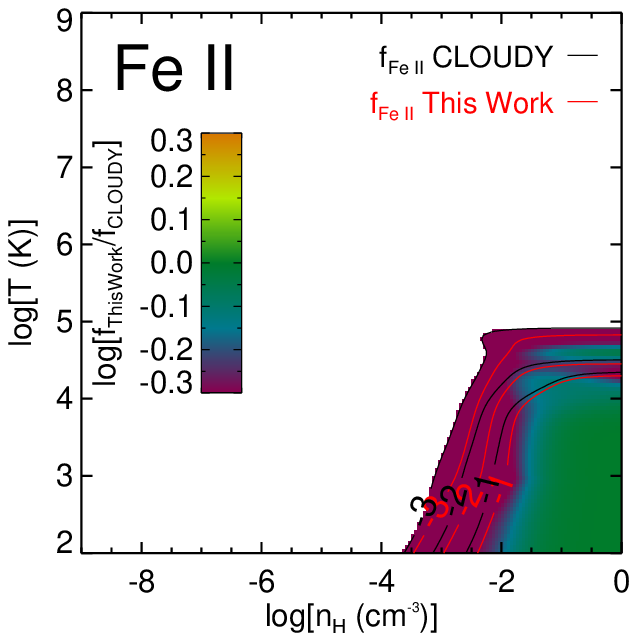}}
  \subfigure{\setlength{\epsfxsize}{0.32\textwidth}\epsfbox{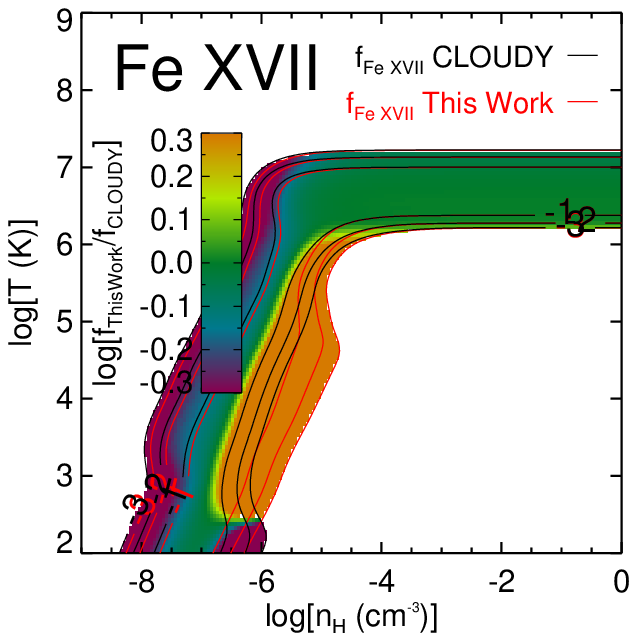}}
\caption{A selection of equilibrium abundances of common ions compared
  between our method (red contours) and CLOUDY (black contours).  The
  colour bar shows the ratio of the ion fractions predicted by the two
  methods.  In both cases, we turn on Auger ionisation and charge
  transfer.  In general the agreement is very good as can be seen by
  comparing the black and red contours, although differences are
  highlighted by the very sensitive colour scaling that we choose.
  Where photo-ionisation is dominant for $\HI$ and $\HeII$, the
  agreement is excellent, but small differences occur in the regime
  where collisional ionisation dominates and $\HI$ becomes primarily
  neutral; this has to do with our assumption of case A recombination,
  which breaks down in this regime.  In general, Li-like and He-like
  ions ($\CIV$, $\OVII$, and $\NeVIII$) show good agreement although
  slight differences can exist at higher photo-ionised densities where
  $f_{x_{i}}$ drops off precipitously with increasing density.
  $\MgII$ shows good agreement except where $\HI$ fractions are
  underestimated, and the electron density is thus different between
  the two cases, which affects low ions like $\MgII$ in the regime
  where collisional ionisation becomes important.  We also show some
  species in the bottom row that are more spurious, likely owing to a
  more complex set of radiative and dielectronic recombination
  coefficients used in CLOUDY ($\SIII$, $\FeII$, and $\FeXVII$).}
\label{fig:comp_noneq_cloudy}
\end{figure*}

We compare equilibrium ionisation fractions calculated using our
method assuming a $z=1$ HM01 background to those of CLOUDY ver. 10.00
for a range of diffuse gas densities and temperatures in Figure
\ref{fig:comp_noneq_cloudy}.  Red and black contours indicate ion
fractions for our method and CLOUDY, respectively.  Black contours
overlap red contours in most cases, indicating excellent agreement.
The shading indicates the ratio log[$f_{\rm This Work}/f_{\rm
    CLOUDY}$] where either ion fraction is $>10^{-3}$ if a metal ion
or $>10^{-8}$ if H or He; green indicates a good fit, while purple
(orange) indicates our method yields too low (high) $f_{x_i}$ compared
to CLOUDY.  Note that our shading is very sensitive to differences
between the two methods in order to highlight slight inconsistencies,
and often the contours nearly overlap when shading is off-green.

For primordial species, the two methods agree well, except for $\HI$
at $T\approx$15,000 K and $\nh > 0.1 \cmc$ where collisional
ionisation causes a rapid transition in temperature between neutral
and highly ionised.  We assume case A recombination in our method
aimed to track diffuse ionised gases, and case B recombination is more
applicable when the gas is optically thick to Lyman series photons.
Nonetheless the two sets of contours still nearly overlap.

$\OVII$ and $\NeVIII$ show excellent agreement, except at higher
photo-ionised densities where $f_{x_i}$ drops off rapidly with
increasing density.  This is also apparent in Figure
\ref{fig:ioncloudycheck} and is of little practical concern.  $\CIV$
also shows excellent agreement, although slight deviations occur where
collisions dominate below $T=10^5$ K, which is also of little
concern since $f_{\rm CIV}$ is very low here.  $\MgII$ shows good
agreement, but reflects the disagreement with $\HI$ at the same range
of densities around $T=15,000$ K, owing to spurious $\HI$ and electron
densities.  Again, the contours show excellent agreement.  The bottom
three panels of Figure \ref{fig:comp_noneq_cloudy} show a selection of
some of the worst agreeing species, which likely owe to more complex
treatment of radiative and dielectronic recombination in CLOUDY
ver. 10.00.  These species are also less studied in laboratory
experiments and the relevant atomic data is therefore often less well
constrained.  The excellent agreement we demonstrate between cooling
efficiencies calculated by CLOUDY and us indicates that these ion
abundance differences are not important for the dynamics of gas, but
they may be important for the diagnostics of the gas if these
particular species are observed.  Comparisons with CLOUDY for
additional ions can be found on the website
http://noneq.strw.leidenuniv.nl.

\begin{figure*}
  \subfigure{\setlength{\epsfxsize}{0.32\textwidth}\epsfbox{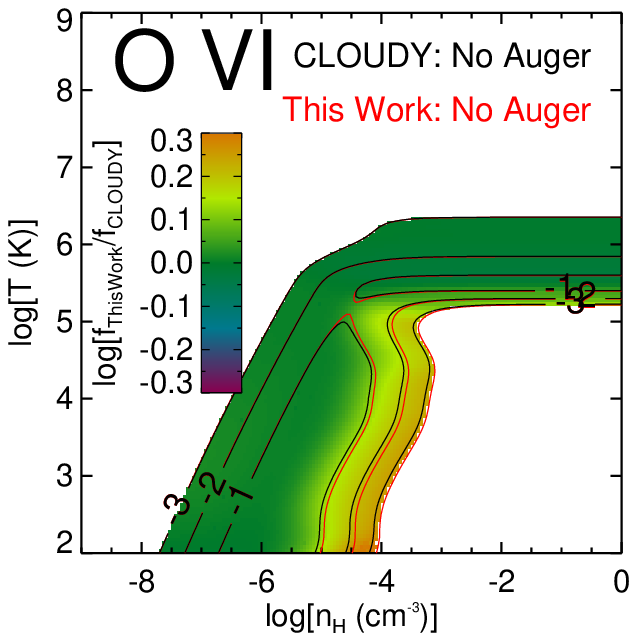}}
  \subfigure{\setlength{\epsfxsize}{0.32\textwidth}\epsfbox{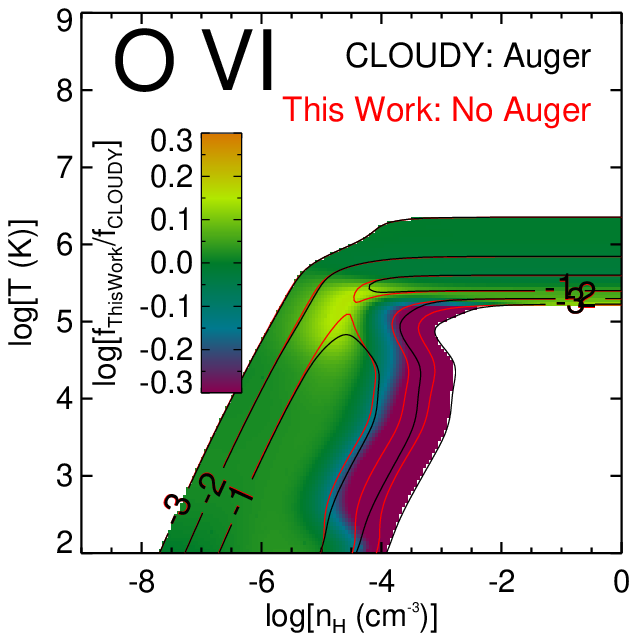}}
  \subfigure{\setlength{\epsfxsize}{0.32\textwidth}\epsfbox{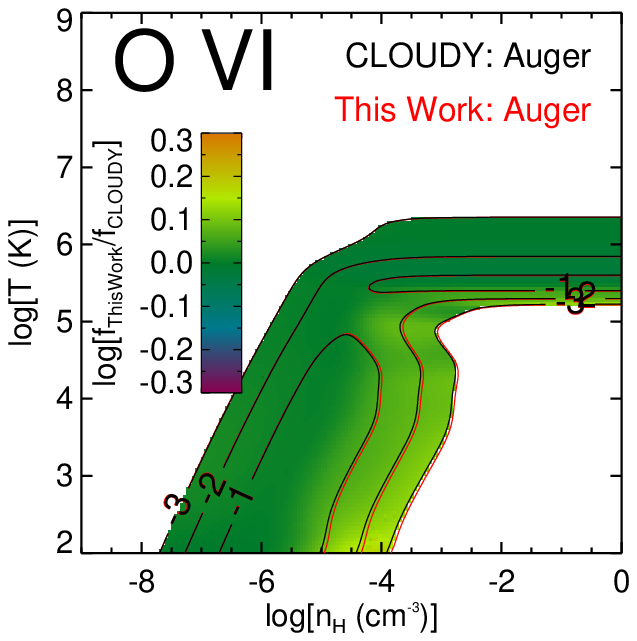}}
\caption{Comparison of $\OVI$ fractions predicted using our method
  calculated in equilibrium (red contours) and CLOUDY (black
  contours).  Contours correspond to logarithmic ionisation fractions,
  and colour shading indicates the ratio of our method's ionisation
  fraction over that of CLOUDY, according to the logarithmic colour
  bar.  From left to right, we show the $\OVI$ fractions without Auger
  ionisation in either case (left), with Auger ionisation on for
  CLOUDY only (centre), and with Auger ionisation on in both methods
  (right).  Auger ionisation boosts the $\OVI$ fraction at
  photo-ionised temperatures at the high density end.}
\label{fig:comp_OVI}
\end{figure*}

In Figure \ref{fig:comp_OVI} we demonstrate the effect of Auger
ionisation on $\OVI$ at $z=1$.  In the left panel we have Auger turned
off both in CLOUDY and our method; the centre panel has Auger turned
on in CLOUDY only; and the right panel shows it turned on in both.  In
the case of oxygen, photo-ionisation of the 1s shell of $\OI$ through
$\OV$ leads most often to the removal of two electrons, which is why
$\OVI$ is more ionised with Auger ionisation turned on, especially at
photo-ionised temperatures at densities above the peak of the
photo-ionised $\OVI$ fraction.  Hence, Auger ionisation is important
for $\OVI$ strengths at these densities and temperatures, and would be
about 2$\times$ lower if Auger ionisation is ignored.

\begin{figure*}
  \subfigure{\setlength{\epsfxsize}{0.32\textwidth}\epsfbox{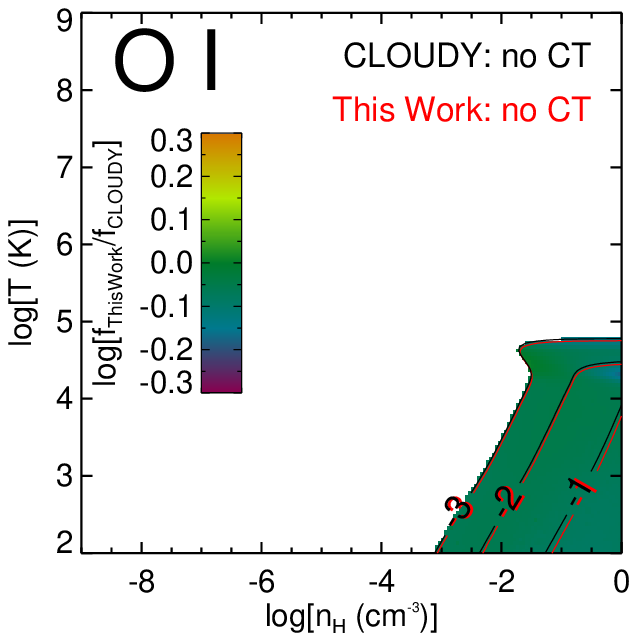}}
  \subfigure{\setlength{\epsfxsize}{0.32\textwidth}\epsfbox{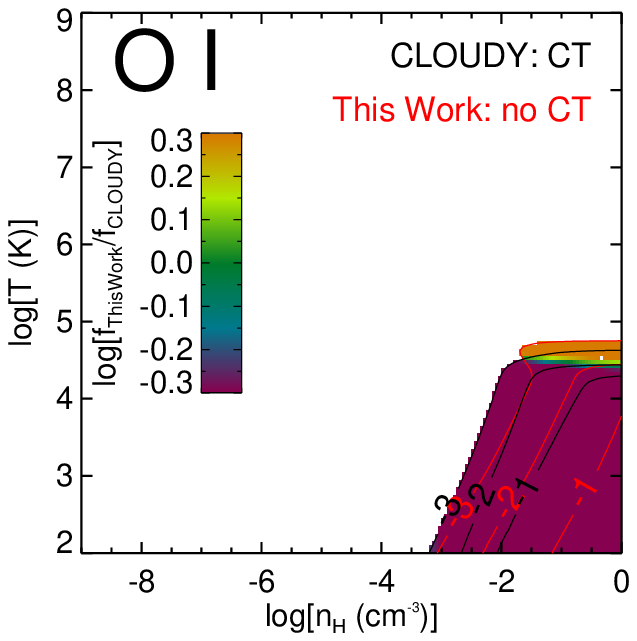}}
  \subfigure{\setlength{\epsfxsize}{0.32\textwidth}\epsfbox{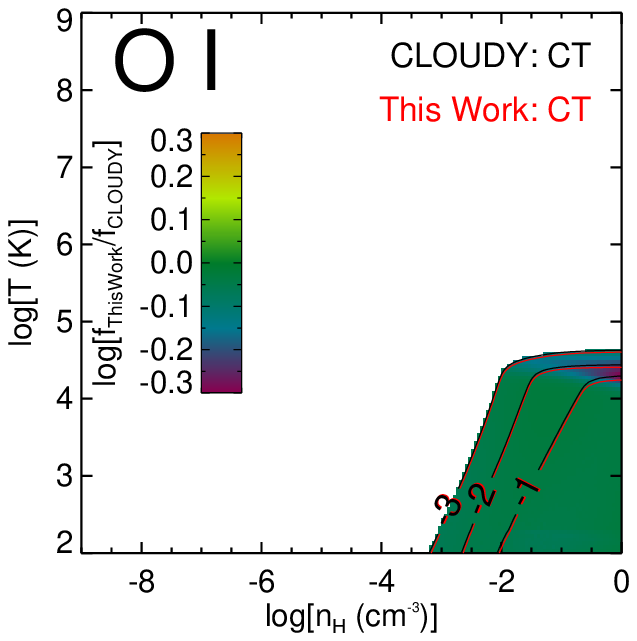}}
\caption{A similar plot as Figure \ref{fig:comp_OVI}, but for $\OI$
  and the effect of charge exchange.  Red contours indicate our
  ionisation fractions, and black contours indicate CLOUDY results.
  From left to right, we show the $\OI$ fractions without charge
  transfer in either case (left), with charge transfer on for CLOUDY
  only (centre), and with charge transfer on in both cases (right).
  Charge transfer is very important for $\OI$ owing to its similar
  ionisation potential with $\HI$, and it is also significant for many
  of the other lower ions.}
\label{fig:comp_OI}
\end{figure*}

We perform an analogous comparison for $\OI$ and the effect of charge
transfer (CT) in Figure \ref{fig:comp_OI} as we did for $\OVI$ and the
Auger effect in Figure \ref{fig:comp_OVI}: from left to right the
panels show no CT in either case, CT only in CLOUDY, and CT in both.
If CT is not turned on, the $\OI$ fraction is underestimated by an
order of magnitude at temperatures where it is found.  $\OI$ shows
extreme sensitivity to charge transfer owing to $\OI$ having a very
similar ionisation potential as $\HI$.

\end{document}